\documentclass{pasj01}
\input{pasjsup.sty}
\usepackage[switch,mathlines]{lineno}
\begin{document}

\author{Mariko \textsc{Kimura}\altaffilmark{1,2,*}
        Hitoshi \textsc{Negoro}\altaffilmark{3}, 
        Shinya \textsc{Yamada}\altaffilmark{4},
        Wataru \textsc{Iwakiri}\altaffilmark{5},
        Shigeyuki \textsc{Sako}\altaffilmark{6,7},
        Ryou \textsc{Ohsawa}\altaffilmark{8}
}

\email{mariko-kimura@se.kanazawa-u.ac.jp}

\altaffiltext{1}{Advanced Research Center for Space Science and Technology, Colledge of Science and Engineering, Kanazawa University, Kakuma, Kanazawa, Ishikawa 920-1192}

\altaffiltext{2}{Cluster for Pioneering Research, Institute of Physical and Chemical Research (RIKEN), 2-1 Hirosawa, Wako, Saitama 351-0198}

\altaffiltext{3}{Department of Physics, Nihon University, 1-8 Kanda-Surugadai, Chiyoda-ku, Tokyo 101-8308}

\altaffiltext{4}{Department of Physics, Rikkyo University, 3-34-1 Nishi-Ikebukuro, Toshima-ku, Tokyo 171-8501}

\altaffiltext{5}{International Center for Hadron Astrophysics, 
Chiba University, Chiba 263-8522}

\altaffiltext{6}{Institute of Astronomy, Graduate School of Science, The University of Tokyo, 2-21-1 Osawa, Mitaka, Tokyo 181-0015}

\altaffiltext{7}{The Collaborative Research Organization for Space Science and Technology, The University of Tokyo, 7-3-1Hongo, Bunkyo-ku, Tokyo 113-0033}

\altaffiltext{8}{National Astronomical Observatory of Japan, 2-21-1 Osawa, Mitaka, Tokyo 181-8588}

\title{
Evolution of X-ray and optical rapid variability during the low/hard state in the 2018 outburst of MAXI J1820$+$070 $=$ ASASSN-18ey
}

\Received{} \Accepted{}


\SetRunningHead{Kimura et al.}{Shot analyses of MAXI J1820$+$070}

\maketitle

\begin{abstract}

We performed shot analyses of X-ray and optical sub-second flares observed during the low/hard state of the 2018 outburst in MAXI J1820$+$070.
\textcolor{black}{Optical shots were less spread than X-ray shots}.
The amplitude of X-ray shots was the highest at the onset of the outburst, and they faded at the transition to the intermediate state.
The timescale of shots was $\sim$0.2~s, and we detected the abrupt spectral hardening synchronized with this steep flaring event.
The time evolution of optical shots was not similar to that of X-ray shots.
These results suggest that accreting gas blobs triggered a series of magnetic reconnections at the hot inner accretion flow in the vicinity of the black hole, which enhanced X-ray emission and generated flaring events.
The rapid X-ray spectral hardening would be caused by this kind of magnetic activity.
Also, the synchrotron emission not only at the hot flow but also at the jet plasma would contribute to the optical rapid variability.
We also found that the low/hard state exhibited six different phases in the hardness-intensity diagram and the correlation plot between the optical flux and the X-ray hardness.
The amplitude and duration of X-ray shots varied in synchrony with these phases.
This time variation may provide key information about the evolution of the hot flow, the low-temperature outer disk, and the jet-emitting plasma.

\end{abstract}

%\linenumbers
%\pagewiselinenumbers

\section{Introduction} \label{sec:intro}

Rapid stochastic variability with timescales of subseconds is one of the most intriguing observational features in accreting stellar-mass black-hole systems.
Low-mass X-ray binaries (LMXBs) are close binary systems composed of a black hole or a neutron star (the primary star) and a low-mass star (the secondary star).
The transferred gas from the Roche-lobe-filling secondary star forms an accretion disk around the primary star.
Transient LMXBs exhibit sudden brightening of the accretion disk (e.g., \cite{tan96XNreview} and \cite{che97BHXN} for reviews of observational properties), which is called an outburst due to thermal-viscous instability \citep{min89BHXN,las01DIDNXT}.
The system becomes much brighter during outbursts over broad wavelengths, and sub-second light variability is easily observed.
A lot of multi-wavelength observations of sub-second flares in LMXBs and related theoretical works have been performed for a long time (e.g., \cite{kan01j1118varcorrelation,gan10gx339,pai19j1357,vel11sscmodel,man96ADAFfluctuation,pou99magnetic,mac03globalMHD}).
Despite extensive research, the radiation process and the mechanism for launching this variability remain subjects of intense debate.

ASASSN-18ey was discovered in optical by the All-Sky Automated Survey for SuperNovae (ASSAS-SN) project \citep{ASASSN} on 2018 March 6 \citep{tuc18a18ey}.
After this detection, the Monitor of All-sky X-ray Image (MAXI: \cite{mat09MAXI}) found an X-ray transient MAXI J1820$+$070 on 2018 March 12 \citep{kaw18atel11399}, which was identified with ASASSN-18ey via optical follow-up observations \citep{den18atel11400}.
Since this object became exceedingly brighter during this outburst because of its small distance ($\sim$3~kpc; \cite{gan19gaiadistance,art20j1820}), and a huge amount of follow-up observations carried out over broad wavelengths showed the black-hole nature of this system \citep{bag18a18ey,shi19j1820}.
The black-hole mass ($M_1$) and the secondary star's mass ($M_2$) were estimated to be $\sim$8.5~$M_{\odot}$ and $\sim$0.6~$M_{\odot}$, respectively, by optical spectroscopy and photometry \citep{tor20j1820,nii21j1820}.
The emission from the accretion disk or jet ejections overshined that from the secondary star at optical wavelengths during the 2018 outburst \citep{shi18j1820}.

Black-hole LMXBs show different kinds of spectral states during outburst, and sub-second stochastic variability is observable during the low/hard state \citep{miy92BHCvariation}.
It is considered that, in this state, a hot, optically-thin, and geometrically-thick accretion flow exists in the vicinity of the black hole, and the low-temperature, optically-thick, and geometrically-thin accretion disk is truncated far from the black hole.
Typically, the inner edge of the disk approaches towards the central black hole with the increase of the system luminosity.
The system enters the high/soft state in which the radiation from the multi-temperature disk is dominant over that from the hot flow.
In the state transition from the low/hard state to the high/soft state, the system may pass through the intermediate state with both strong disk emission and hot flow emission (\cite{don07XB} for a comprehensive review).
High-velocity and collimated plasma ejections called jets are probably launched during the low/hard state and the intermediate state \citep{fen04grs1915}.
While the above picture is the classical one, recent observations suggest that the inner edge of the accretion disk is close to the innermost stable circular orbit (ISCO) even in the low/hard state in some black-hole LMXBs. 
The threshold of the mass accretion rate for triggering the evaporation of the hot flow is still in debate \citep{mil06j1753,ram07j1753,rei10hardstate,gar15gx339}.
MAXI J1820$+$070 stayed in the low/hard state during $\sim$100~d from the onset of the outburst \citep{shi19j1820}, providing the best opportunity for investigating the detailed property of the rapid variability.
%the multi-wavelength spectrum and variability changed with time even in this state.

Power spectrum analyses and time-lag analyses have been performed so far for multi-wavelength light curves obtained during the 2018 outburst in MAXI J1820$+$070 \citep{pai19j1820,bui19j1820,xu20j1820,axe21j1820,wan20j1820,wan21j1820,pal21afpi,kaw22j1820,tho22j1820,pra22j1820,oma23j1820}.
However, the properties of sub-second flares such as the amplitude, the duration, and the light-curve shape were not well discussed.
We observed this object during the 2018 outburst by Tomo-e Gozen, an optical wide-field video observation system \citep{sak18a18ey}.
Also, this object was frequently observed by the X-ray telescope NICER onboard the International Space Station (ISS) \citep{NICER}.
Since the absolute time accuracy of Tomo-e Gozen and NICER is better than 1~ms, they are suitable for observing rapid variability in LMXBs.
%The absolute time tagging accuracy is better than 300 ns.
%This telescope is, therefore, suitable for observing rapid X-ray variability in LMXBs.
We carried out shot analyses of optical and X-ray sub-second flares by using Tomo-e Gozen and NICER data taken during the low/hard state of the 2018 outburst in MAXI J1820$+$070.
Shot analyses in which averaged profiles of sub-second flares are generated were proposed by \citet{neg94cygx1shot}.
The advantage of this method is that the average flare shape can be preserved.

Our aim in this paper is to investigate the entire evolution of the properties of optical and X-ray sub-second flares during the 2018 outburst in MAXI J1820$+$070, to compare the optical and X-ray properties, and to discuss the main characteristics of them by referring to other works.
This paper is structured as follows.
Section 2 describes the observation and data reduction. 
Section 3 explains our analysis method and gives the results of optical and X-ray shot analyses. 
In section 4, we interpret the extracted shot properties and compare them with the results reported by other observational and theoretical works.
We give a summary in section 5.

\section{Observations and analyses} \label{sec:observation}

\subsection{Tomo-e Gozen}

Tomo-e Gozen is an optical wide-field video observation system composed of 84 chips of CMOS image sensors on the 1.05 m Kiso Schmidt telescope \citep{sak18tomoe}.
This is capable of obtaining consecutive frames with timestamps of 0.2-ms absolute accuracy.
The frame rate of Tomo-e Gozen is increased by reducing the fields-of-view of the sensors.
We observed MAXI J1820$+$070 with about 67.1 fps in a partial readout mode, where the field of view size was 142.56 arcsec by 71.28 arcsec.
The observation log is given in Table E1 in the supplementary information.
We performed relative photometry with a neighbor reference star (TYC 444-2244-1) \citep{gaia23}.
The fluxes of both stars were measured by the PSF photometry with Gaussian kernels.
We confirmed that time variations of measured flux due to atmospheric fluctuations were negligible in comparison with the variability of this source by the relative photometry method with bright sources in the same frame.
The error of the aperture photometry is composed of the photon noise, the atmospheric fluctuations, the readout noise, and the current noise.
Since MAXI J1820$+$070 was bright during the outburst, the photon noise was dominant.
However, the relative photometry did not work if a thin cloud had passed through the field of view.
Then, the brightness fluctuated.
We removed this kind of bad data for analyses by the criterion that the flux error divided by the flux is higher than 0.4.
All observation times were converted to barycentric Julian date (BJD).\footnote{The BJD is the Julian Date corrected for differences in the Earth's position with respect to the barycenter of the solar system. We use BJD in terrestrial time (TT). The time system TT established by the international astronomical union (IAU) differs from UTC by 32.184 s plus accumulated leap seconds.}

\subsection{NICER}

NICER is composed of 56 pairs of silicon drift detectors (SDDs) with a large effective area in soft X-rays ranging between 0.2 and 12~keV \citep{NICER}.
The absolute time tagging accuracy is better than 300 ns.
NICER monitored MAXI J1820$+$070 during its 2018 outburst, and we here focus on the low/hard state till BJD 2458304.
In this work, we used HEAsoft version 6.32.1 for data reduction and analyses. 
The data were reprocessed with the pipeline tool \texttt{nicerl2}, which used the NICER Calibration Database (CALDB) version later than 2022 October 31 for producing light curves and time-averaged spectra. 
Fifty two modules in NICER's SDDs are operating in orbit, but include two noisy modules, IDs 14 and 34. 
In our data reduction process with \texttt{nicerl2}, we filtered out the data of the noisy two modules. 
The light curves were generated by the tool \texttt{nicer3-lc}.
The time bin of the NICER light curve is 0.015~s, corresponding to the frame rate of the Tomo-e Gozen light curve.
The observation log is given in Table E2 in the supplementary information.
All observation times were converted to barycentric Julian date (BJD) by \texttt{barycorr}.

\section{Results}

\subsection{Correlation between overall X-ray and optical light curves during the low/hard state} \label{sec:lcs}

\begin{figure*}[htb]
\begin{center}
\FigureFile(120mm, 50mm){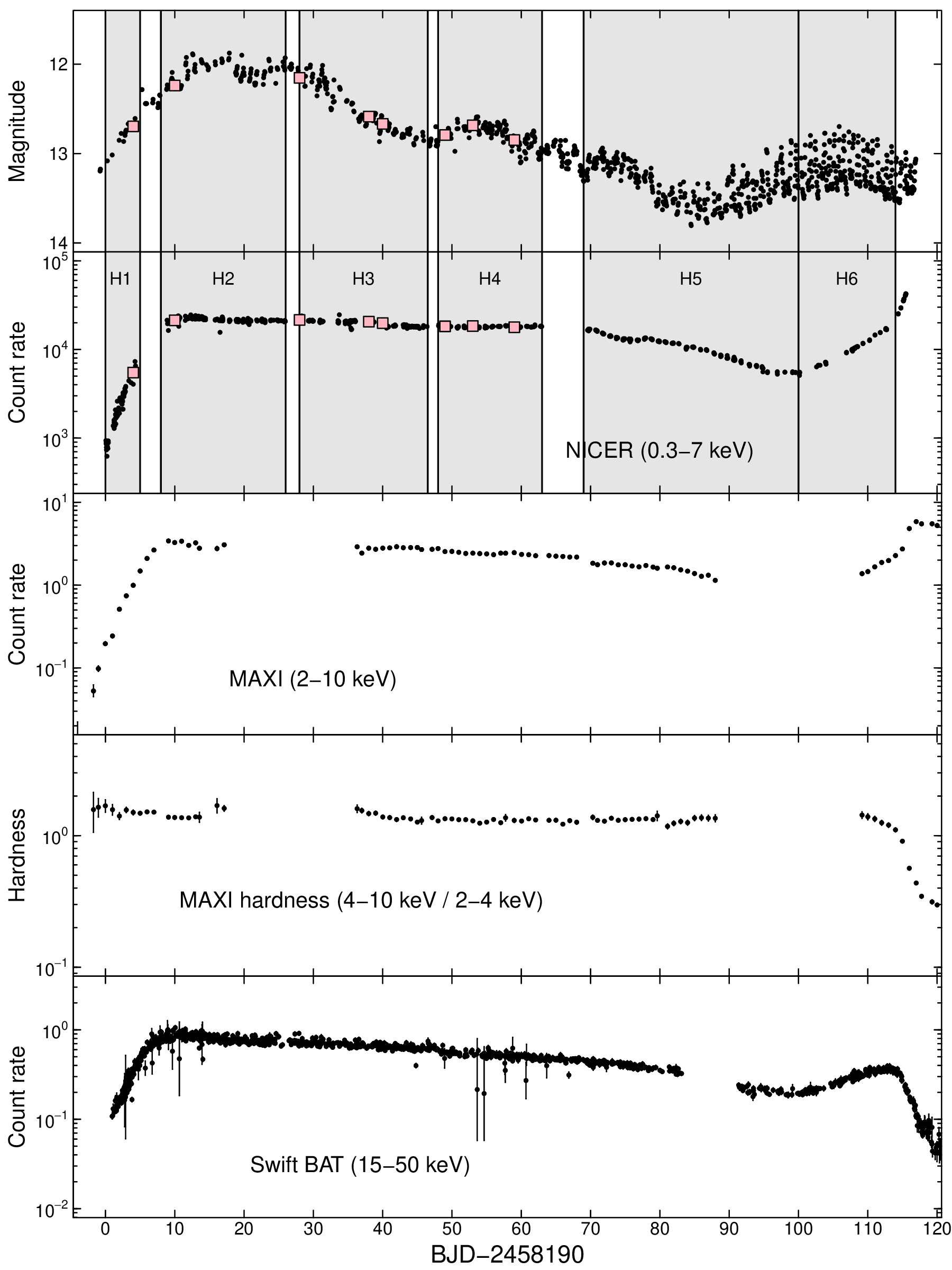}
\end{center}
\caption{
Overall light curves during the low/hard state of the 2018 outburst in MAXI J1820$+$070.
The top panel shows the optical $V$-band light curve provided by the Variable Star Network (VSNET) observers \citep{nii21j1820}.
The second panel shows the NICER 0.3--7~keV light curve.
The third panel represents the MAXI/GSC light curve in the 2--10 keV band.
The fourth panel displays the hardness ratio, which is the ratio of the MAXI 4--10 keV flux to the MAXI 2--4 keV flux.
The bottom panel shows the Swift/BAT light curve in the 15--50 keV band.
The data in the third, fourth, and bottom panels were provided by \citet{shi19j1820}.
We denote by pink rectangles the dates on which semi-simultaneous NICER and Tomo-e Gozen observations were performed in the first and second panels.
Here, BJD 2458190.00000 corresponds to MJD 58189.50045, 2018 March 12 12:00:39 (UTC).
}
\label{overall}
\end{figure*}

We extracted the overall X-ray light curve during the low/hard state from NICER data and compared it with MAXI and Swift/BAT X-ray light curves.
MAXI/GSC data were obtained by the point-spread-function fit method \citep{mor16maxi} using GSC\_2, GSC\_4, GSC\_5, and GSC\_7 camera units.
The Swift/BAT light curve was the same as displayed in \citet{shi19j1820}. 
Figure \ref{overall} represents these light curves and the MAXI 4--10 keV / 2--4 keV hardness ratio.
Incomplete effective-area correction near the edge of the detector in the above method caused artificial sudden or gradual changes in the flux and the hardness ratio before and/or after the data gaps (camera changes) near BJD 2458206, 2458226, and 2458271.
Here, we also plot the optical light curve displayed in \citet{nii21j1820}.
After the release of the NICERDAS software version 11a, the problem of the telemetry saturation\footnote{$<$https://heasarc.gsfc.nasa.gov/docs/nicer/data\_analysis/nicer\_analysis\_tips.html\#Fragmented-Good-Time-Intervals$>$} is addressed.
The baseline of the light curve in the second panel of Figure \ref{overall} is at least not discontinuous.

The rising phase to the outburst maximum continued for $\sim$10~d, and the optical and X-ray flux exponentially decreased after the maximum. 
The optical outburst maximum was delayed to the X-ray outburst maximum by a few days.
The hardness abruptly decreased around BJD 2458305, which represents the X-ray spectral transition.
According to \citet{shi19j1820}, the low/hard state continued till BJD 2458304, and the system entered the intermediate state after that.

\begin{figure*}[htb]
\begin{center}
\begin{minipage}{0.49\hsize}
\FigureFile(80mm, 50mm){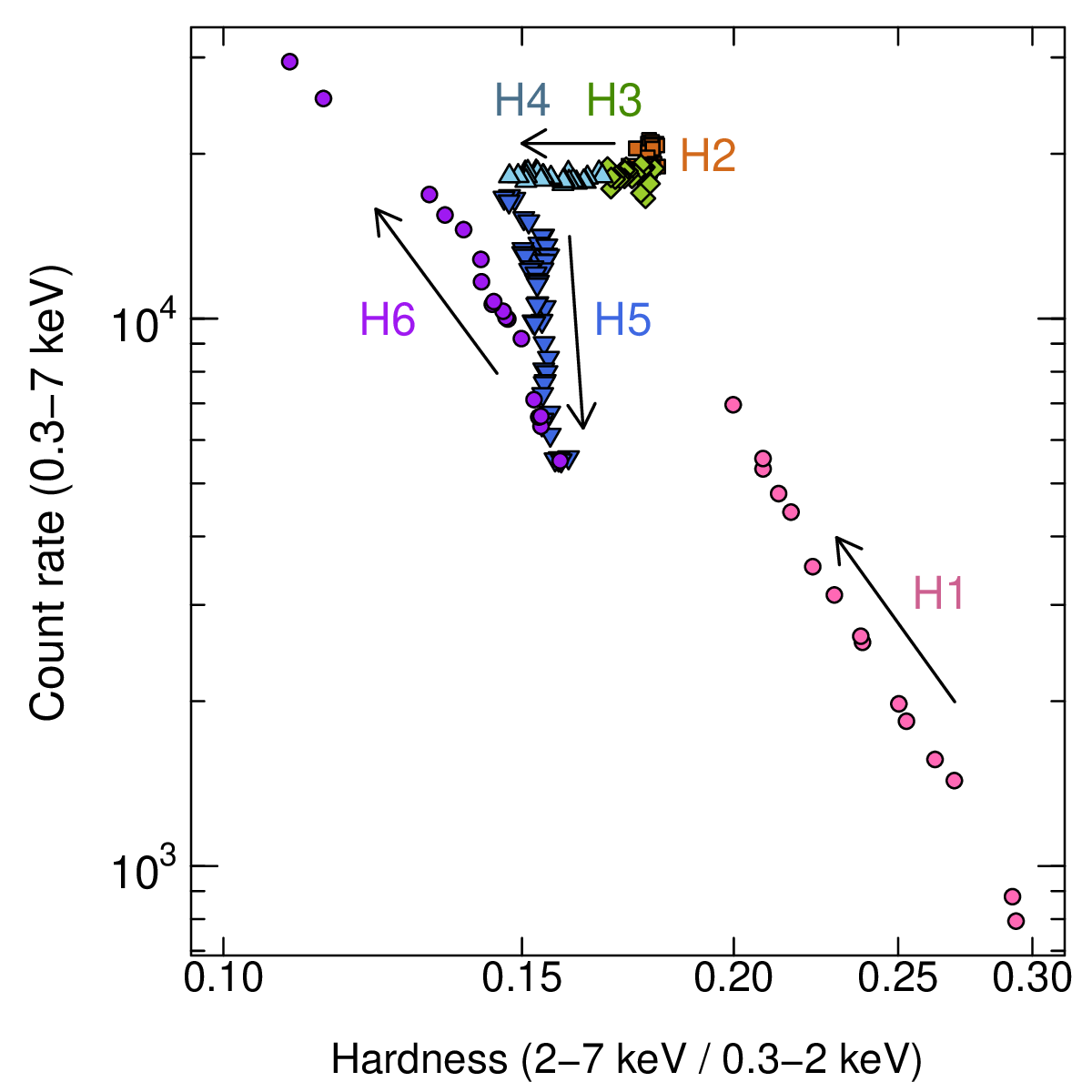}
\end{minipage}
\begin{minipage}{0.49\hsize}
\FigureFile(80mm, 50mm){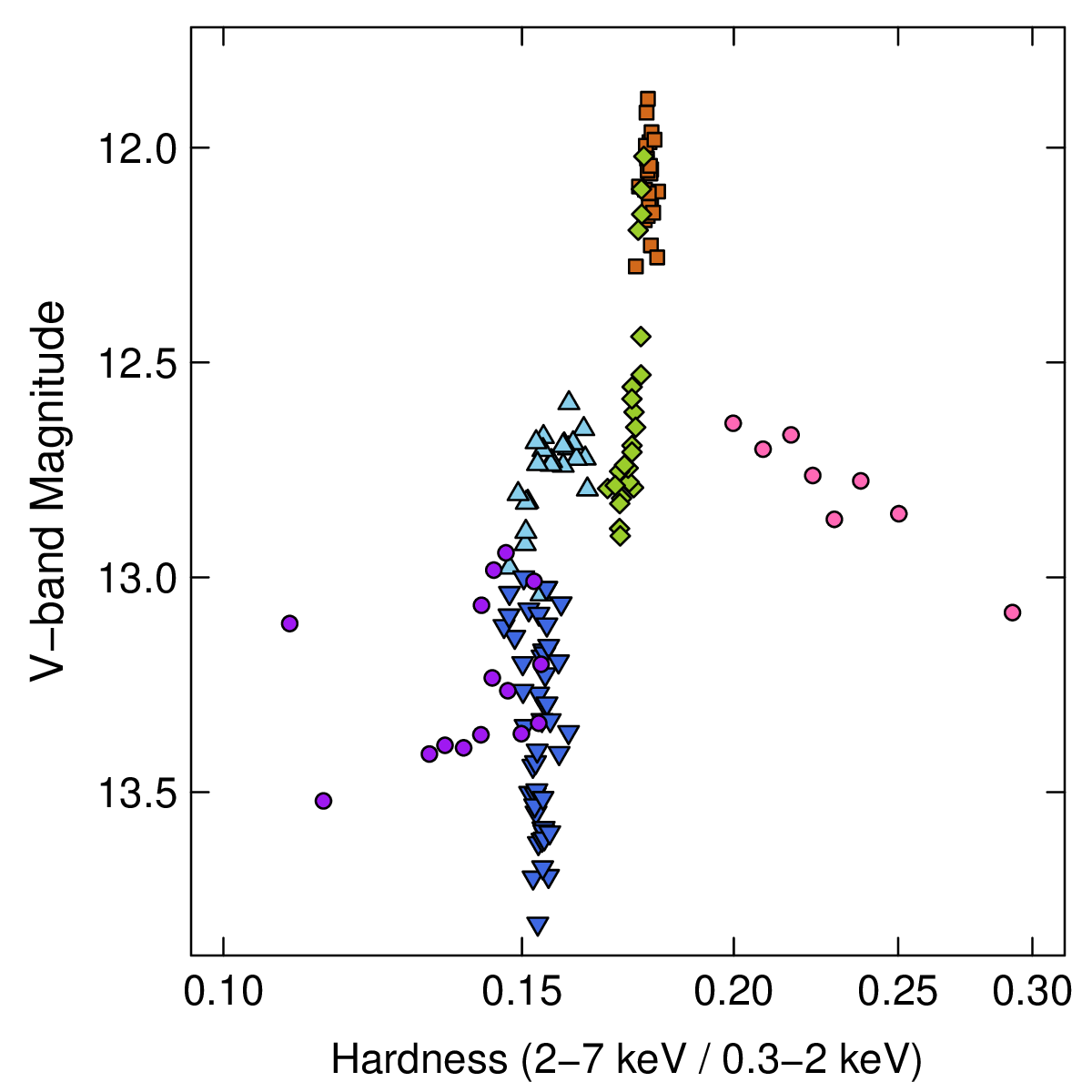}
\end{minipage}
\end{center}
\caption{
(Left) Hardness-intensity diagram with the NICER data of MAXI J1820$+$070 during the low/hard state in the 2018 outburst.
The arrows represent the direction of the time evolution.
The data points denoted with pink, orange, green, light blue, blue, and purple colors represent six different phases named H1, H2, H3, H4, H5, and H6.
(Right) Relation between the optical $V$-band magnitude and the NICER hardness. 
The horizontal axis is the same as that of the left panel.
The data were averaged per day.
The color coding is the same as that in the left panel.
}
\label{q-diagram}
\end{figure*}

We investigated the time evolution of MAXI J1820$+$070 during the low/hard state of the 2018 outburst in the hardness-intensity diagram (see the left panel of Figure \ref{q-diagram}) and the correlation between the NICER hardness ratio and the optical flux (see the right panel of the same figure).
There were six different phases named H1, H2, H3, H4, H5, and H6, which are denoted with different colors. 
The time periods for H1, H2, H3, H4, H5, and H6 are BJD 2458190.0--2458195.0, BJD 2458198.0--2458216.0, BJD 2458218.0--2458236.5, BJD 2458238.0--2458253.0, BJD 2458259.0--2458290.0, and BJD 2458290.0--2458304.0, respectively.
These time periods are represented as grey boxes in the first and second panels of Figure \ref{overall}.
During H1, both the X-ray and optical flux increased towards the outburst maximum, and the system became softer.
During H2, both the X-ray and optical flux showed a plateau but oscillated. The hardness remained at $\sim$0.18.
During H3, the average X-ray flux seemed constant, and the optical flux decreased. The hardness maintained around 0.18 at first and became softer. The boundary between H2 and H3 may be unclear. We set it at the short data gap of the NICER light curve by considering the results of our shot analyses presented in section \ref{sec:xray-shots}. The X-ray shot property changed at the boundary between H2 and H3.
During H4, the hardness became softer, while the X-ray flux seemed constant. The optical flux increased at first and decreased later.
During H5, the X-ray flux suddenly dropped and the hardness slightly increased. The optical flux decreased at first and increased later. The fluctuation of the optical flux became prominent at the latter part.
During H6, the system became brighter in X-rays, and the hardness rapidly decreased. The optical flux fluctuated by $\sim$0.5~mag.
\citet{nii21j1820} reported the optical periodic modulations called superhumps were observed after H5.
The fluctuations of the optical flux during H5 and H6 originated from the superhumps, and the tidal dissipation energy may increase the optical flux in H5 and H6 \citep{osa89suuma}.

The classification of different phases in the low/hard state similar to ours was performed by \citet{pra22j1820} according to the hardness-intensity diagram. Phase II in their classification, in which the central energy of complex iron lines shifted from 6.6~keV to 6.8~keV, was divided into three phases in detail by our investigation.
The boundary between H4 and H5 was pointed out by \citet{wan20j1820} via their X-ray spectral analyses.

\subsection{Shot analyses}

\subsubsection{Methods} \label{sec:method}

\begin{figure*}[htb]
\begin{center}
\begin{minipage}{0.49\hsize}
\FigureFile(80mm, 50mm){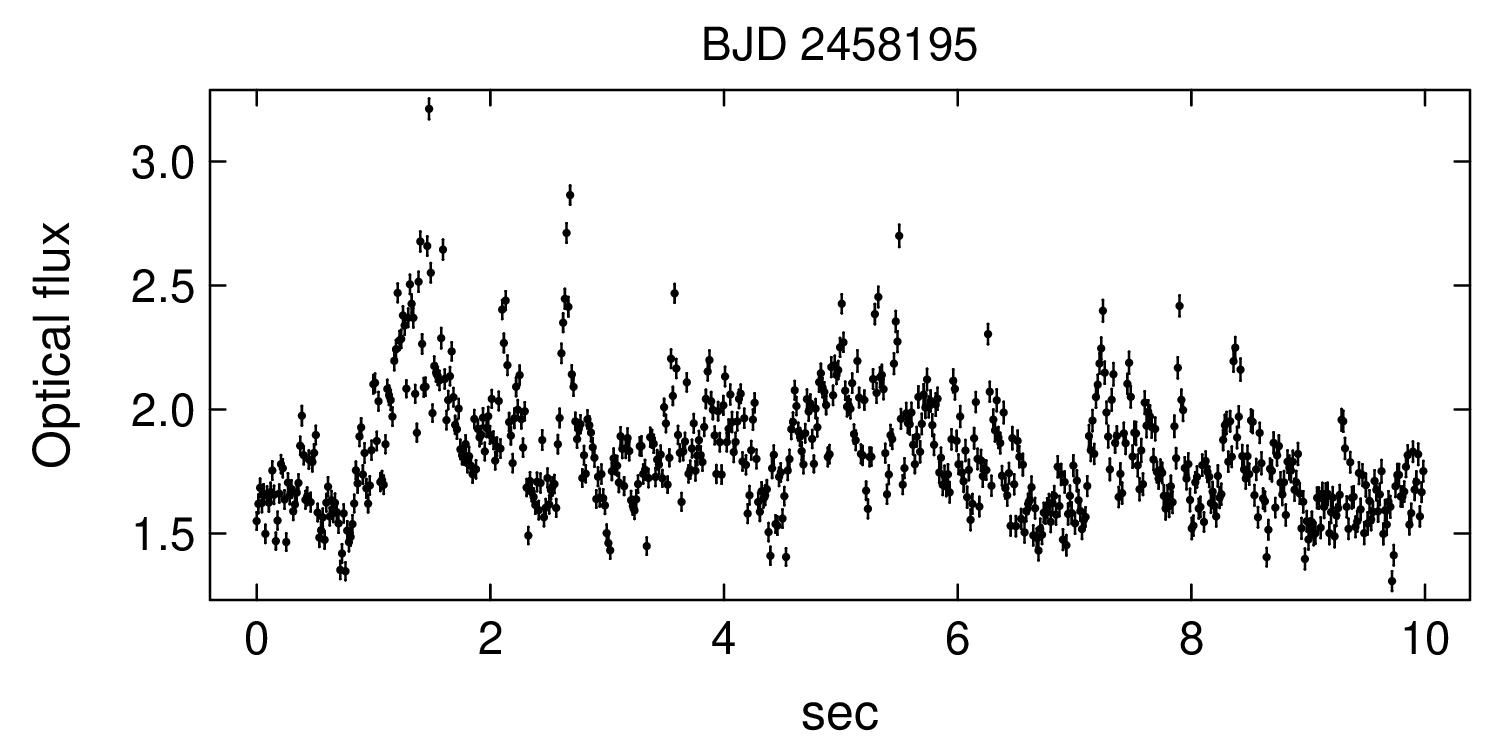}
\end{minipage}
\begin{minipage}{0.49\hsize}
\FigureFile(80mm, 50mm){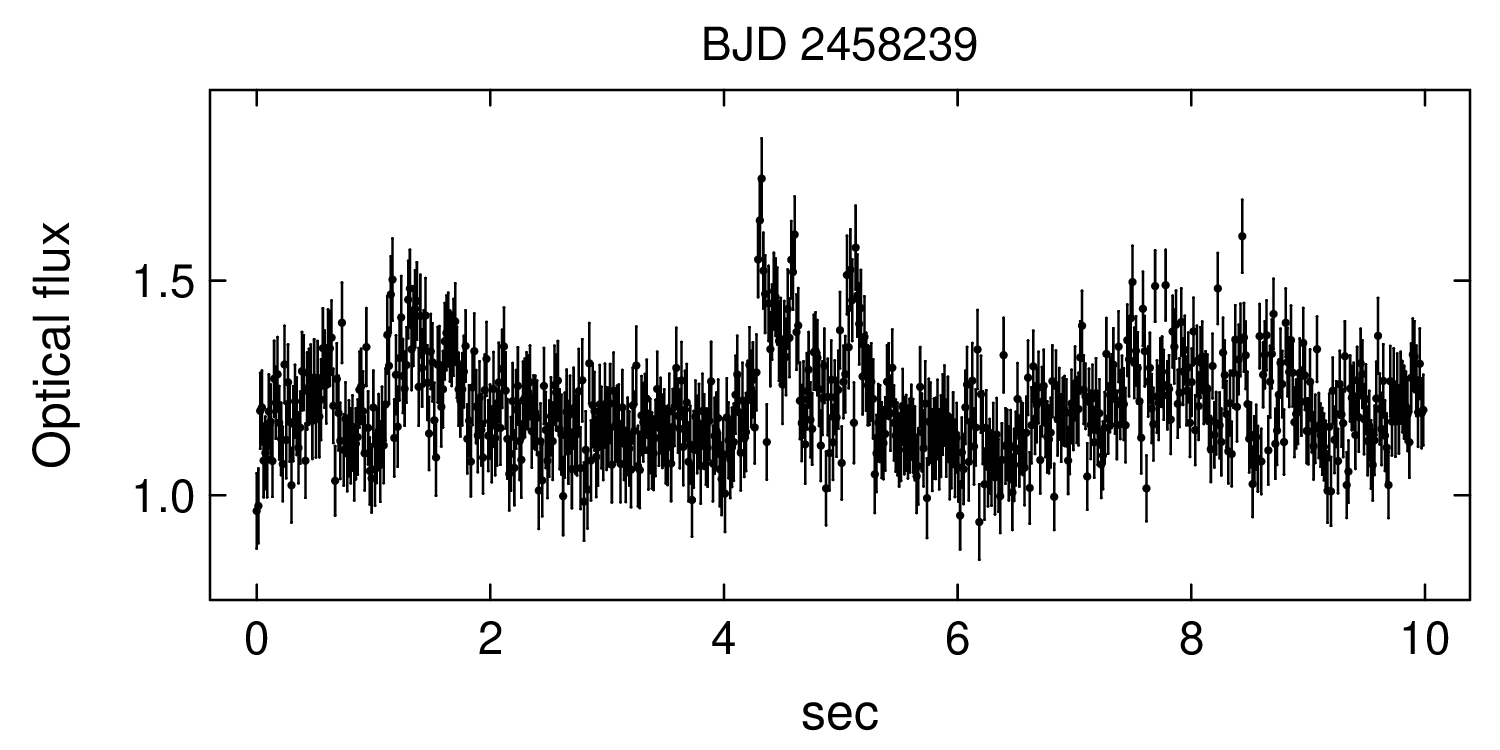}
\end{minipage}
\\
\begin{minipage}{0.49\hsize}
\FigureFile(80mm, 50mm){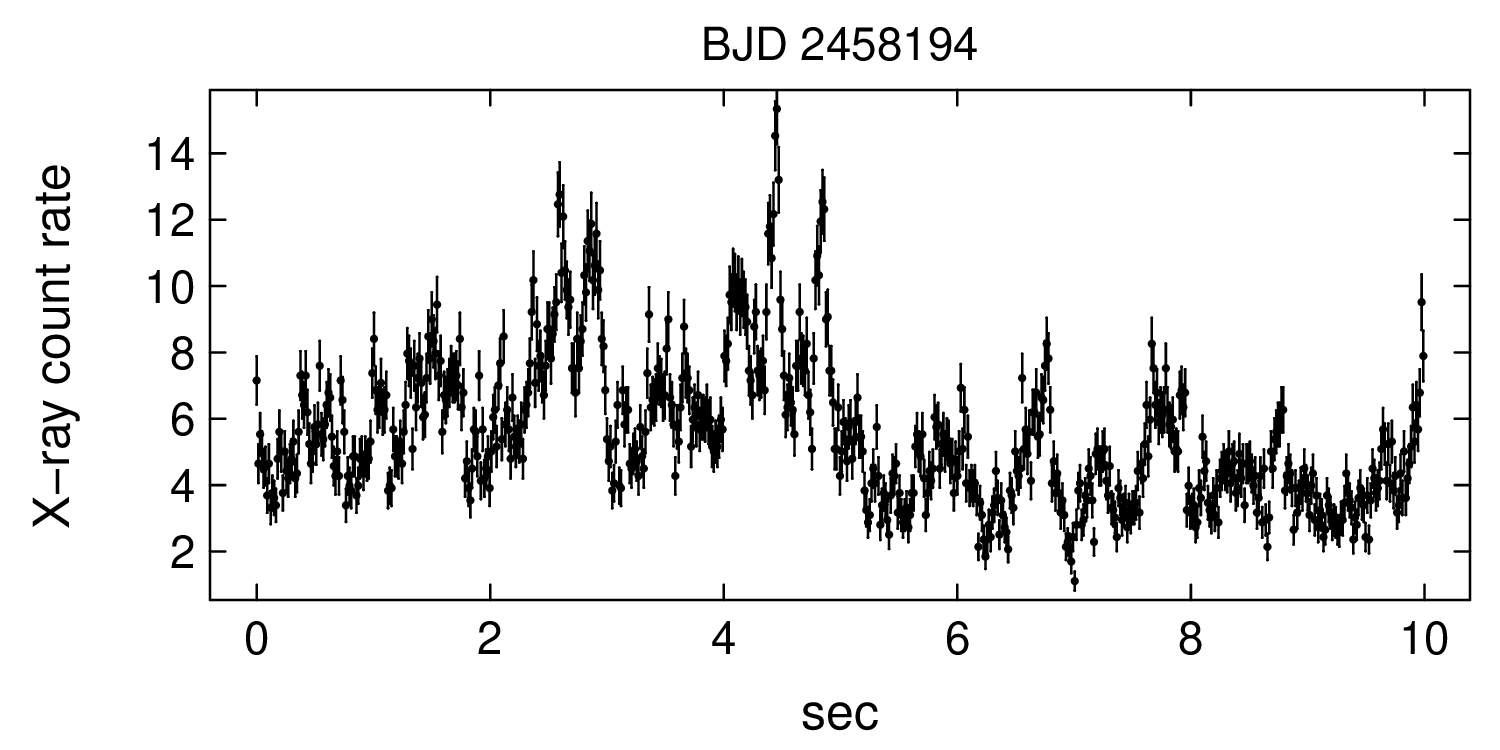}
\end{minipage}
\begin{minipage}{0.49\hsize}
\FigureFile(80mm, 50mm){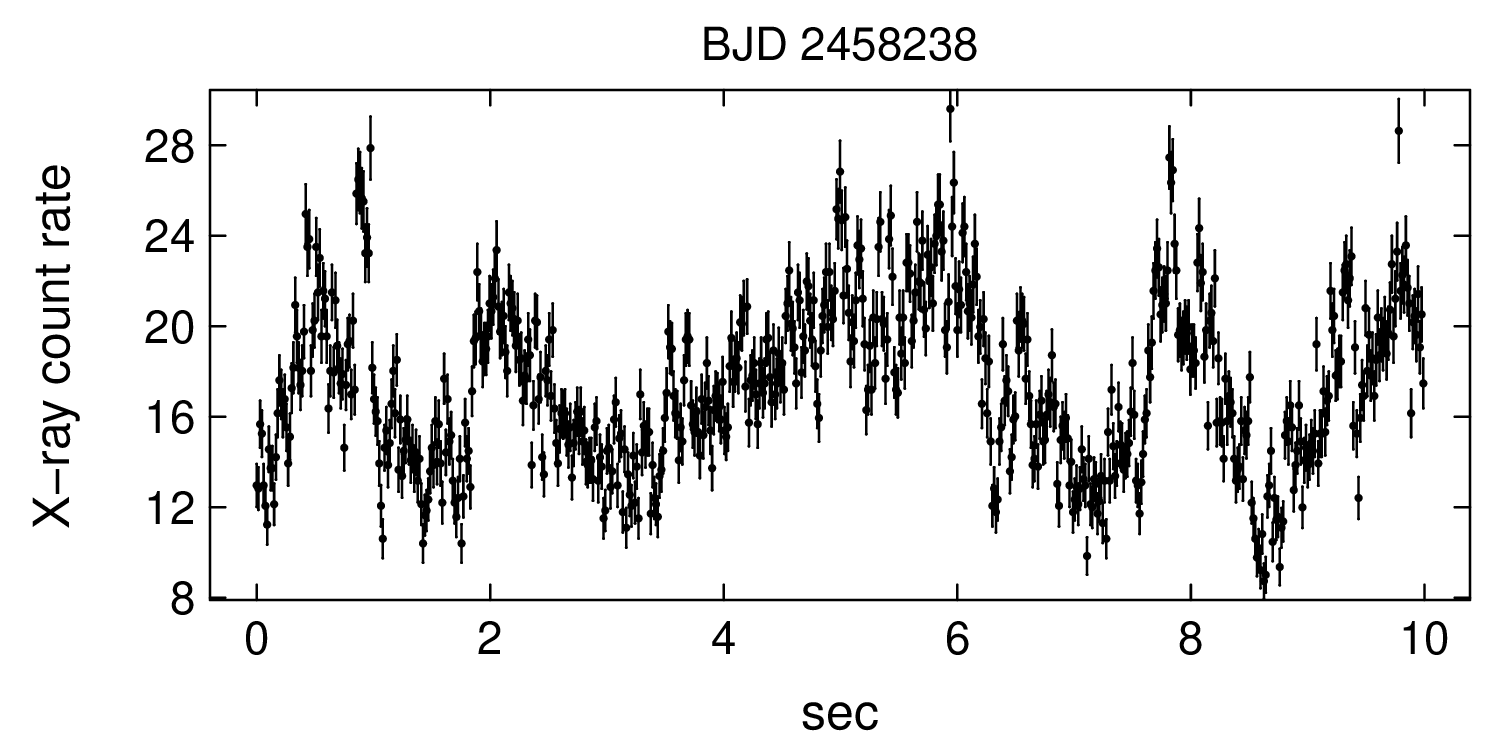}
\end{minipage}
\end{center}
\caption{
A part of original light curves of optical flares observed by Tomo-e Gozen on BJD 2458195 and BJD 2458239 (upper panels) and X-ray flares observed by NICER on BJD 2458194 and BJD 2458238 (lower panels).
The optical flux ($F_{\rm opt}$) in the upper panels is converted from the observed magnitude ($m$) by $F_{\rm opt} = 10^{-m/2.5 + 5}$.
The X-ray count rate in the lower panels is that in the 0.3--7~keV band divided by $10^3$.
}
\label{original-flares}
\end{figure*}

Both of NICER and Tomo-e light curves repeatedly showed sub-second and high-amplitude flares during the low/hard state.
We give some examples of optical and X-ray flares in Figure \ref{original-flares}.
To extract the characteristics of these flares, we performed shot analyses.
This kind of analysis was proposed by \citet{neg94cygx1shot}, and our method is based on the method in this paper.

First of all, we divided light curves per day and estimated the long-term trend of light curves by locally weighted polynomial regression (LOWESS: \cite{LOWESS}).
The parameter $f$ in LOWESS represents the proportion of points that influences the smooth at each value.
We set $f$ at 100~s divided by the total observational length in units of seconds, which varies with the data length.
We also checked the dependence in $f$ of the results by varying $f$ between 0.01--0.1, and confirmed that the shot profile did not change.
Second, we normalized light curves by the extracted long-term trend.
Third, we found the peak time ($t_{\rm p}$) at which the data point has the highest normalized flux over $t_{\rm int}$ before and after, and cut the light curve of $t_{\rm p} - t_{\rm int} < t < t_{\rm p} + t_{\rm int}$, which is called a shot.
We here set $t_{\rm int}$ to 5 s.
Fourth, we superposed all extracted shots per day and took the average normalized flux in each time bin, which makes one average shot profile.
We tried the shot selection with the other two values of $t_{\rm int}$, 2 s and 10 s, and confirmed that $t_{\rm int}$ did not influence the shot profile very much.
The peak flux, however, becomes higher for the longer $t_{\rm int}$ (see the left panel of Figure E1 in the supplementary information).
We hereafter plot the peak flux of the shot profile and the hardness at the peak flux time by open circles as reference values.

Here, we need to be careful not to detect false flares due to Poisson fluctuations.
If the peak flux detected by the above method is due to Poisson fluctuations, the flux of data points near $t_{\rm p}$ is as low as the noise level.
We selected real flares by the criterion that the ratio of the average flux of ten data points before and after $t_{\rm p}$ to the peak flux at $t_{\rm p}$ is higher than 0.3.
We define this ratio as $R_{\rm near}$.
We tried the shot selection with the other two values of $R_{\rm near}$, 0.2 and 0.4, and confirmed that the shot profile was robust (see the right panel of Figure E1 in the supplementary information).

\subsubsection{X-ray and optical semi-simultaneous shots} \label{sec:opt-xray-shots}

\begin{figure*}[htb]
\begin{center}
\begin{minipage}{0.325\hsize}
\FigureFile(50mm, 50mm){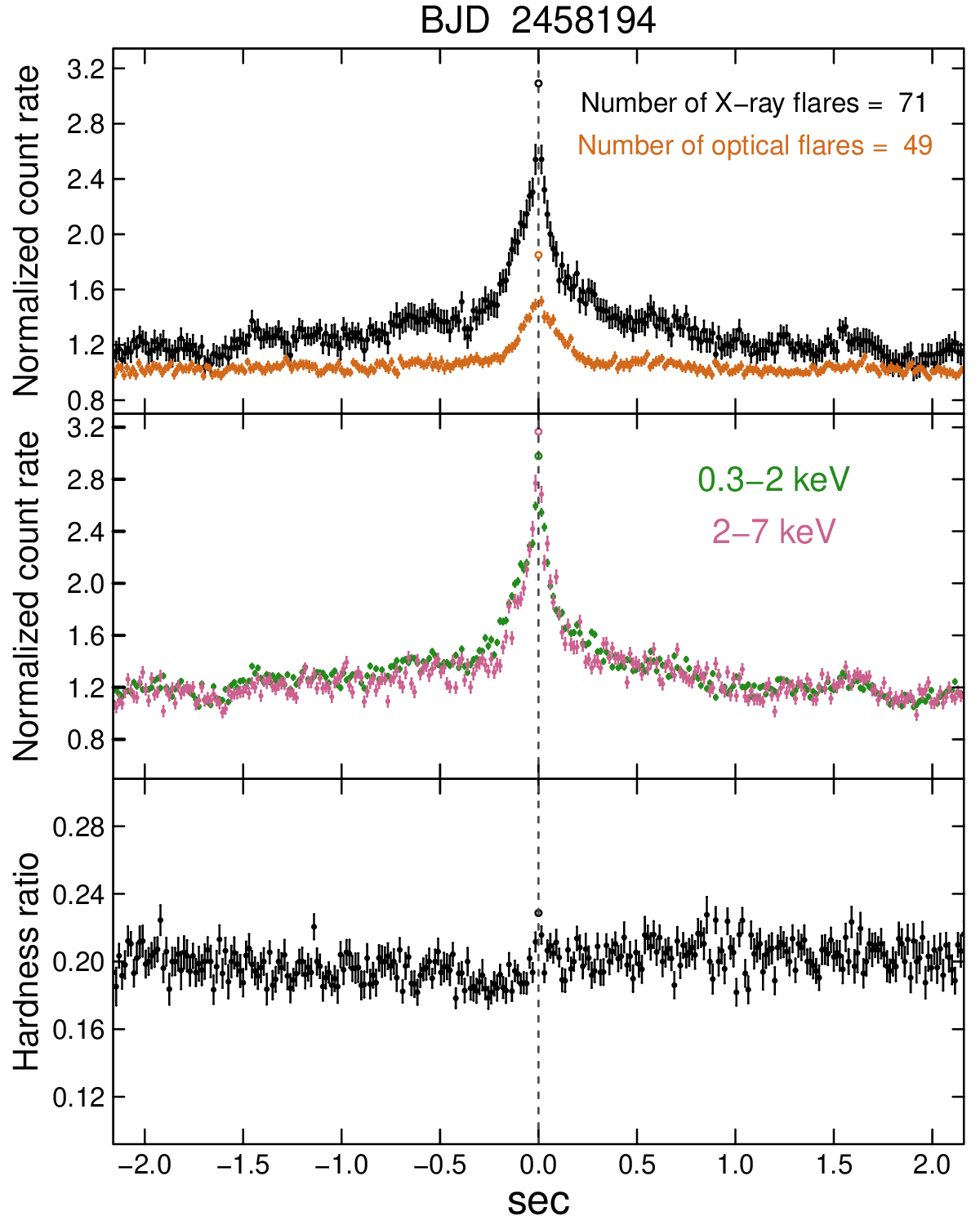}
\end{minipage}
\begin{minipage}{0.325\hsize}
\FigureFile(50mm, 50mm){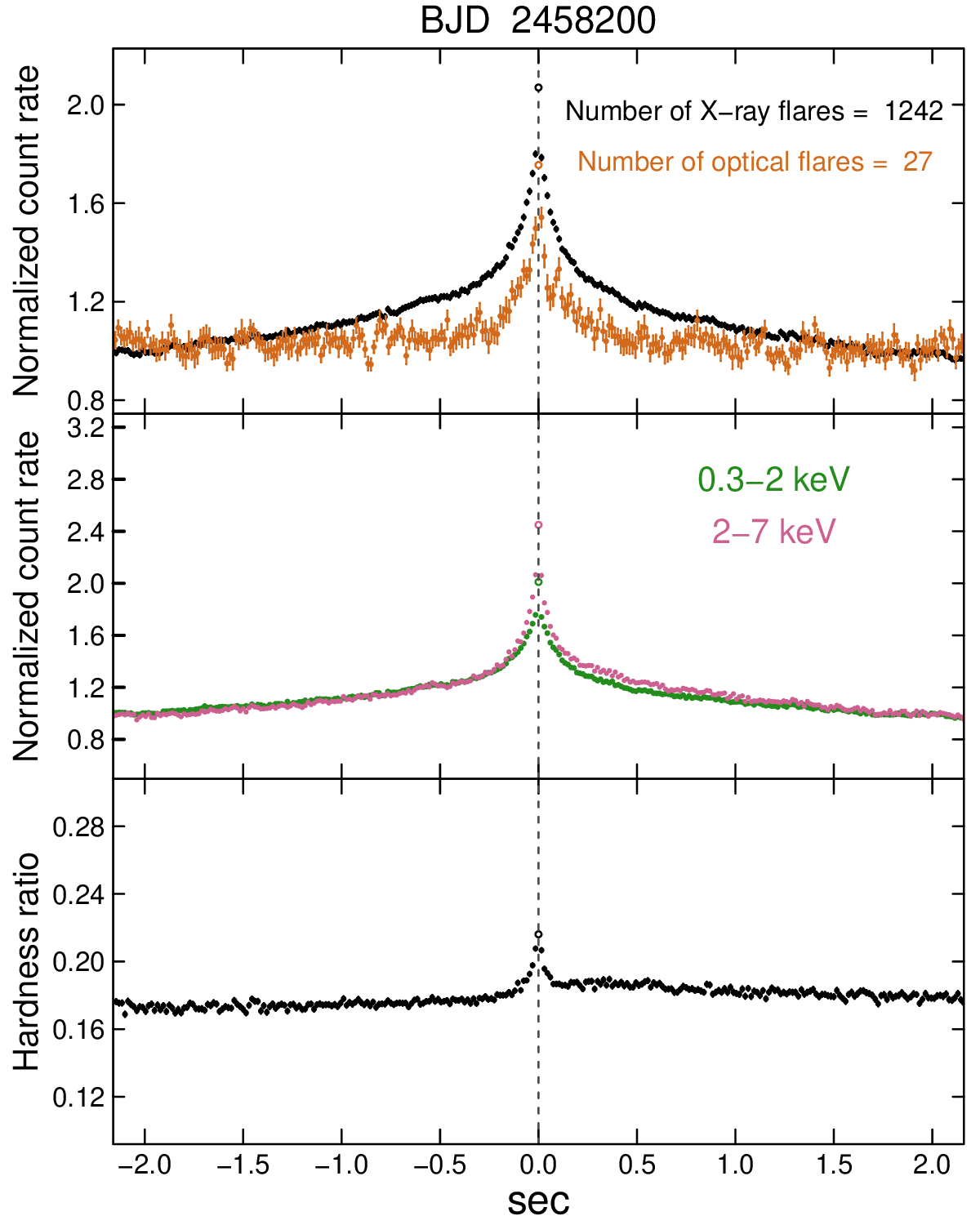}
\end{minipage}
\begin{minipage}{0.325\hsize}
\FigureFile(50mm, 50mm){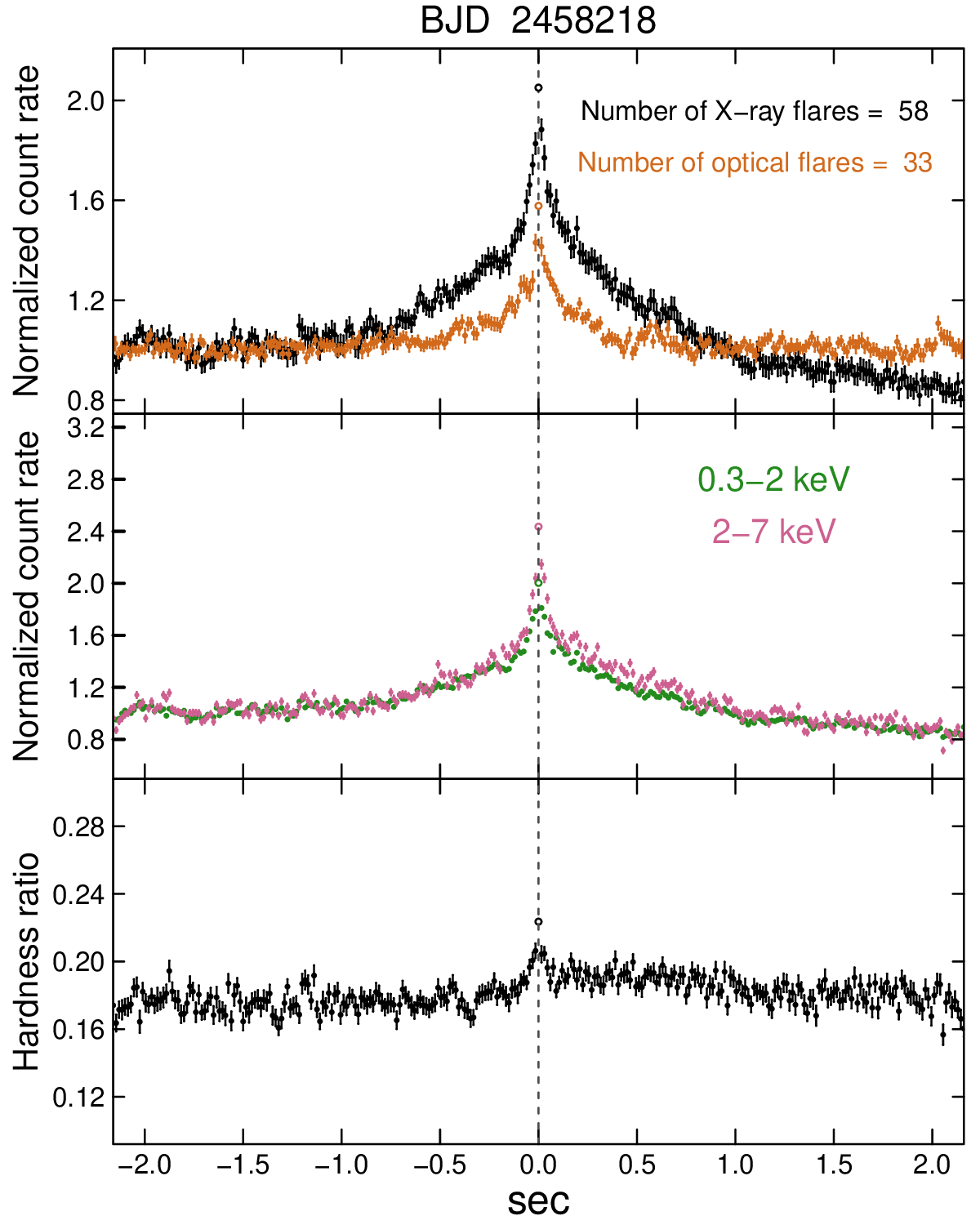}
\end{minipage}
\\
\begin{minipage}{0.325\hsize}
\FigureFile(50mm, 50mm){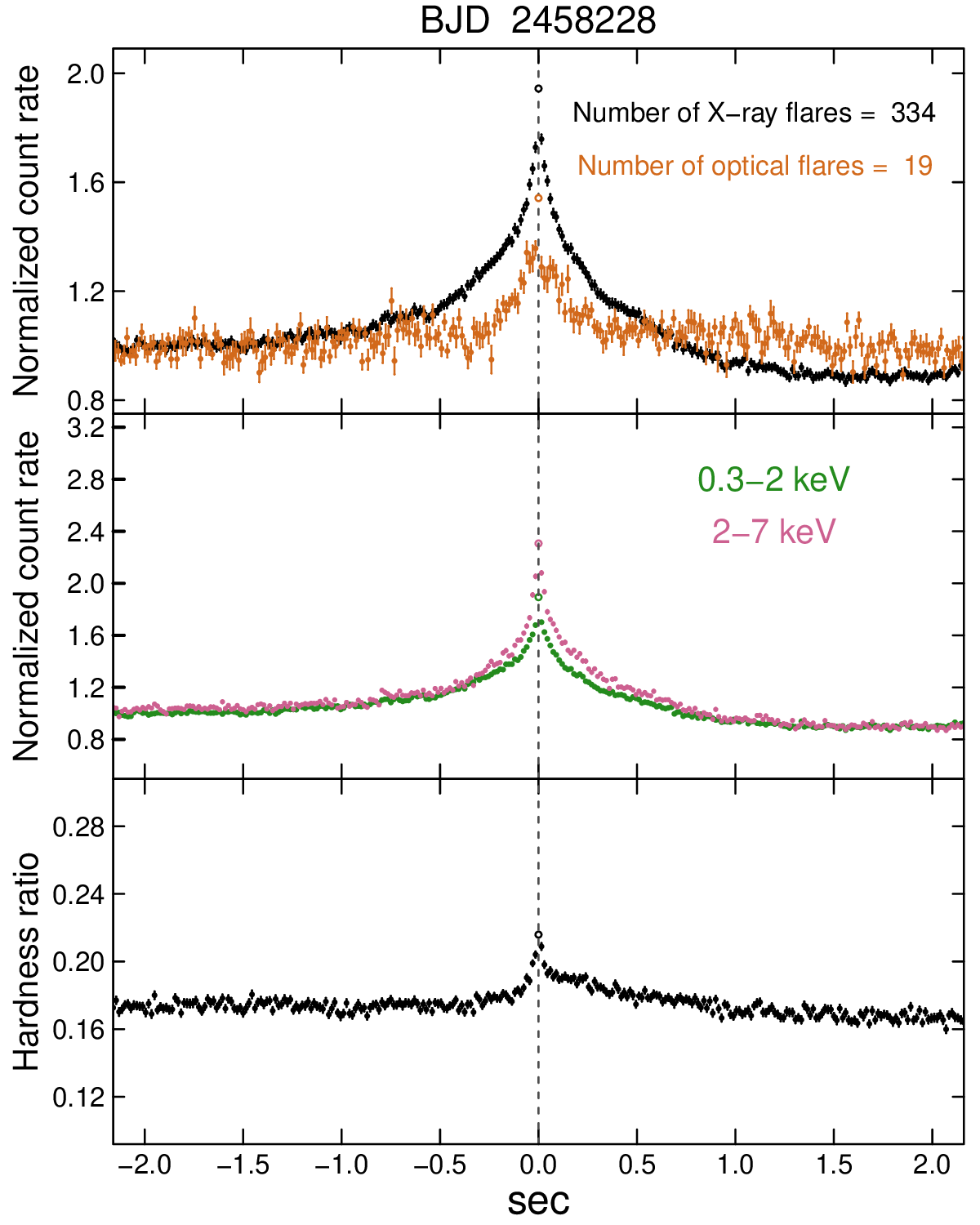}
\end{minipage}
\begin{minipage}{0.325\hsize}
\FigureFile(50mm, 50mm){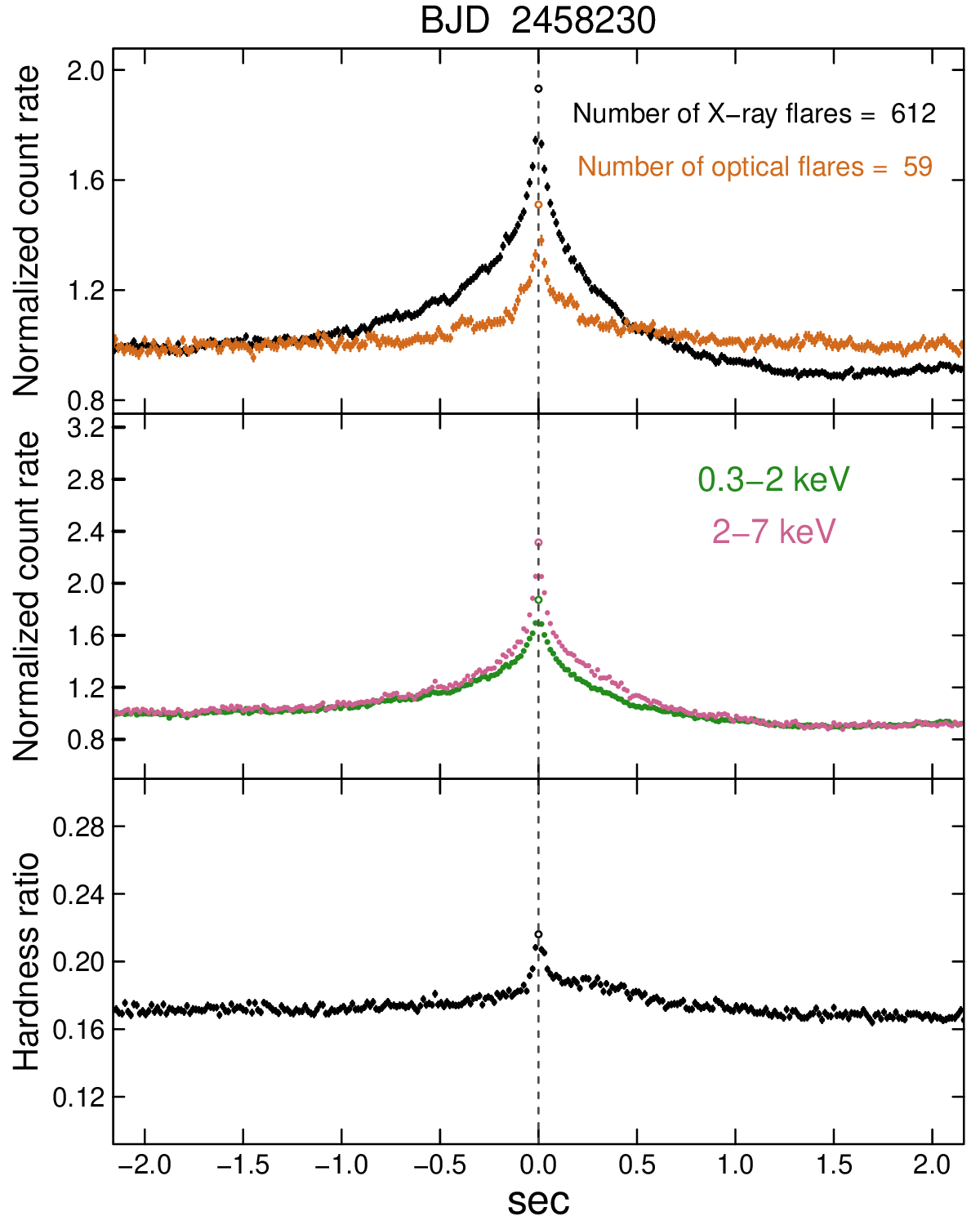}
\end{minipage}
\begin{minipage}{0.325\hsize}
\FigureFile(50mm, 50mm){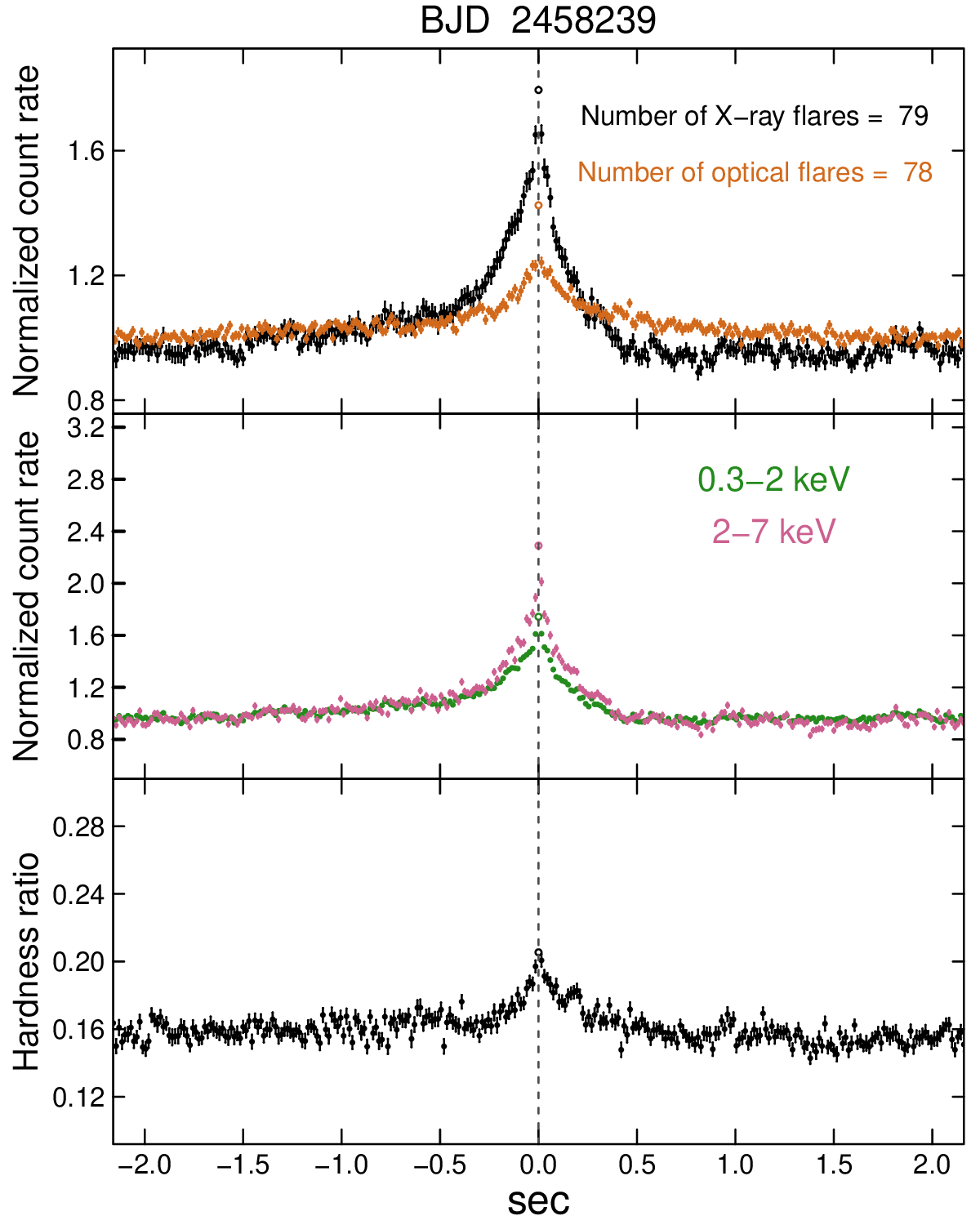}
\end{minipage}
\\
\begin{minipage}{0.325\hsize}
\FigureFile(50mm, 50mm){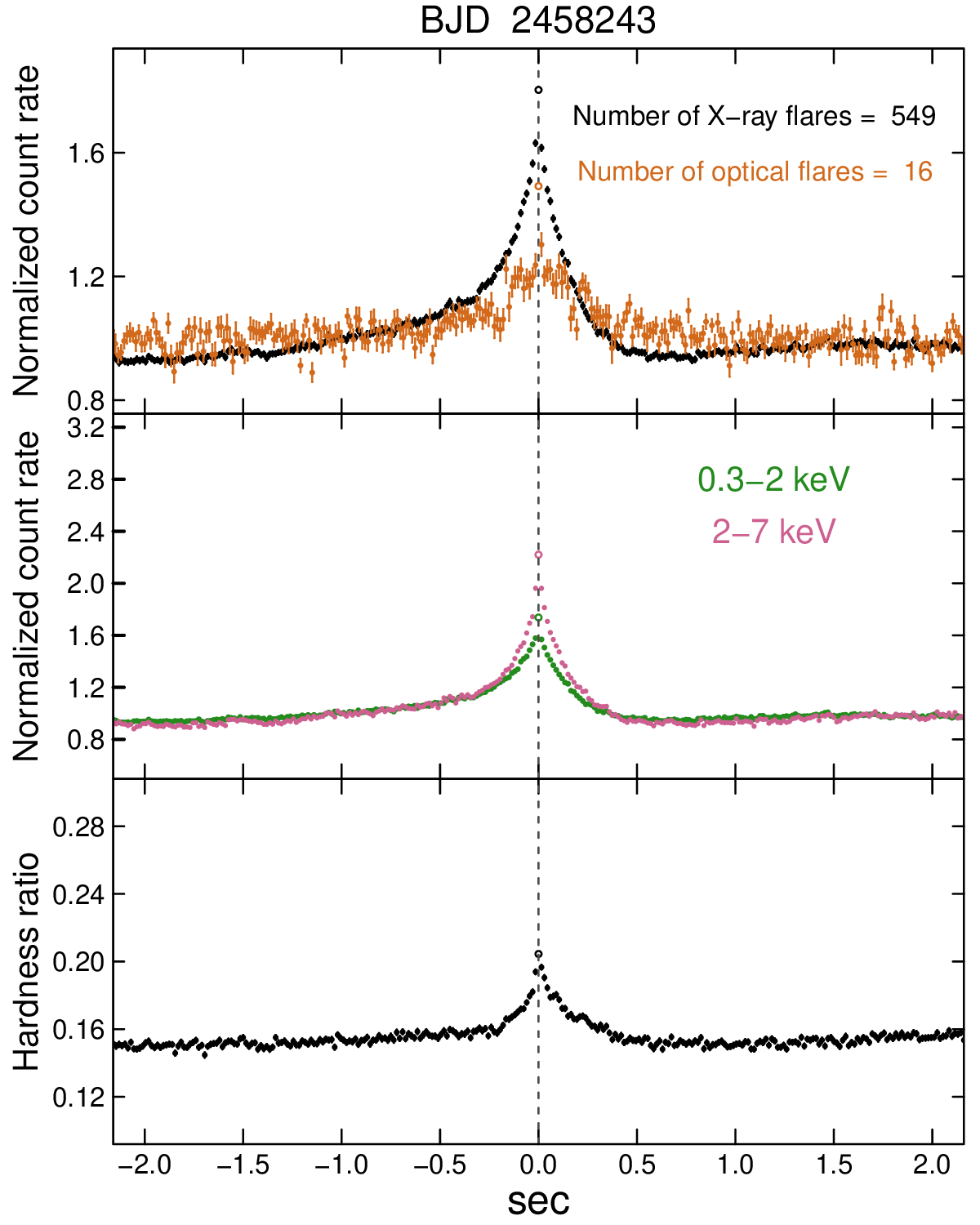}
\end{minipage}
\begin{minipage}{0.325\hsize}
\FigureFile(50mm, 50mm){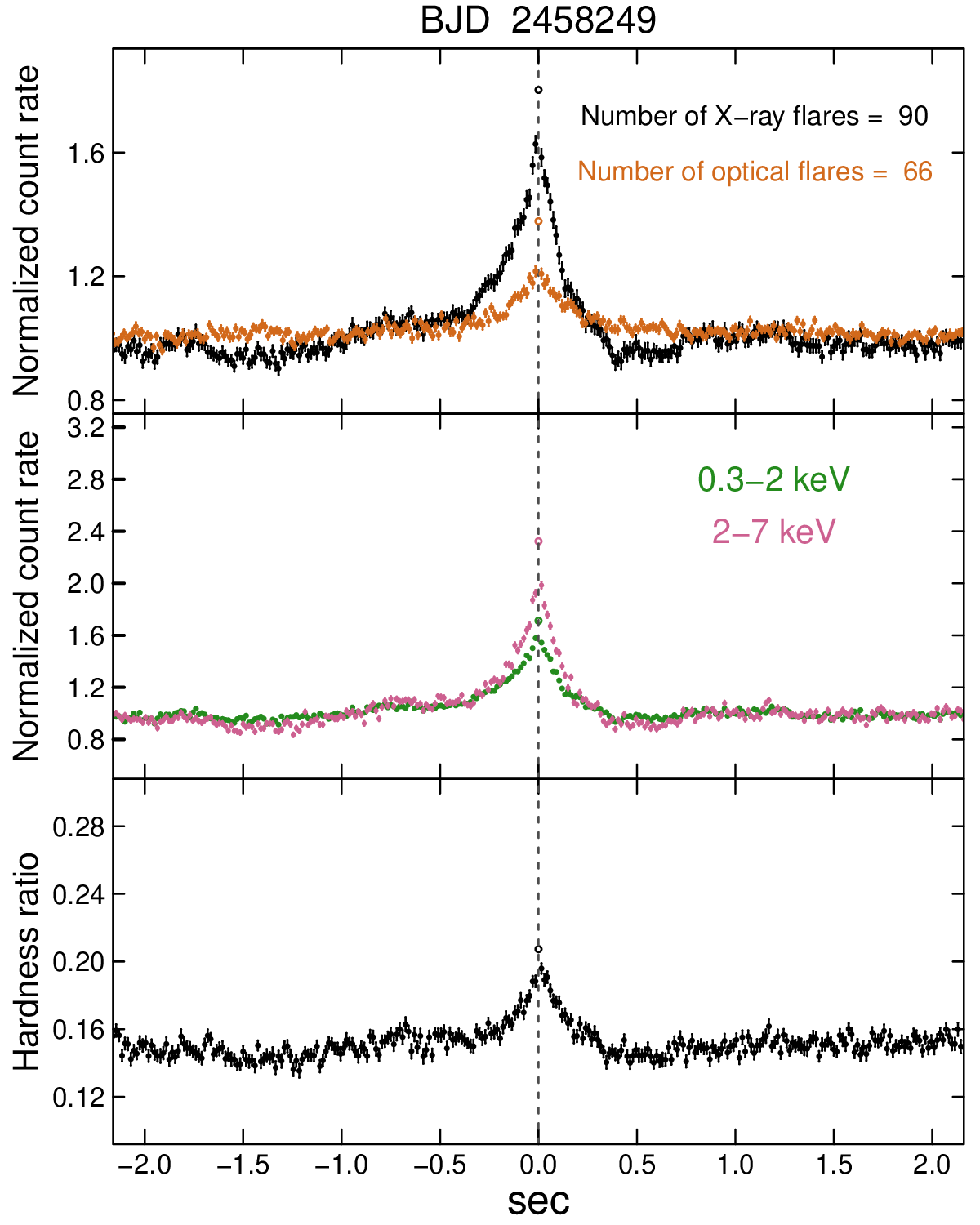}
\end{minipage}
\end{center}
\caption{
Comparison between optical and X-ray shots on the same day during the low/hard state of the 2018 outburst in MAXI J1820$+$070.
The top panel displays the optical shot (orange) and the X-ray shot in the 0.3--7 keV band (black).
The middle panel shows the soft X-ray shot in the 0.3--2 keV band (green) and the hard X-ray shot in the 2--7 keV band (pink).
The bottom panel is the NICER 2--7~keV / 0.3--2~keV hardness ratio of the X-ray shot.
The number of flares superposed in one shot profile is given in the top panel.
}
\label{optical-xray-shots}
\end{figure*}

Several observational dates overlapped between NICER and Tomo-e Gozen, although we failed to obtain completely simultaneous observations because of bad weather.
Figure \ref{optical-xray-shots} represents the average X-ray and optical shots on the same day.
In this figure, we plot the optical and X-ray shots with their peak fluxes aligned.
The amplitude of optical shots was consistently lower than that of X-ray shots.

We estimated the peak flux and duration of each normalized optical and X-ray shot.
We divided the shot profile into the two blocks, the rising part at $t < 0$ and the fading part at $t > 0$, and fitted each of them with the following exponential functions.
\begin{eqnarray}
F(t) &= F_1 \exp (t/\tau_1) + c_1~(t < 0) \nonumber \\ 
F(t) &= F_2 \exp (t/\tau_2) + c_2~(t > 0), 
\label{exp-func}
\end{eqnarray}
where $F_1$ and $F_2$ represent the amplitudes of shots, $\tau_1$ and $\tau_2$ denote the timescales of the rising and the fading parts of shots, and $c_1$ and $c_2$ are the normalized fluxes outside of shots to be determined, respectively.
Here, we omitted the data point at the peak flux.
Since the NICER observations of MAXI J1820$+$070 were performed on other dates except for the dates shown in Figure \ref{optical-xray-shots}, we applied the same fitting for all of the data during the low/hard state.

%\ifnum0=1
\begin{figure*}[htb]
\begin{center}
\begin{minipage}{0.325\hsize}
\FigureFile(50mm, 50mm){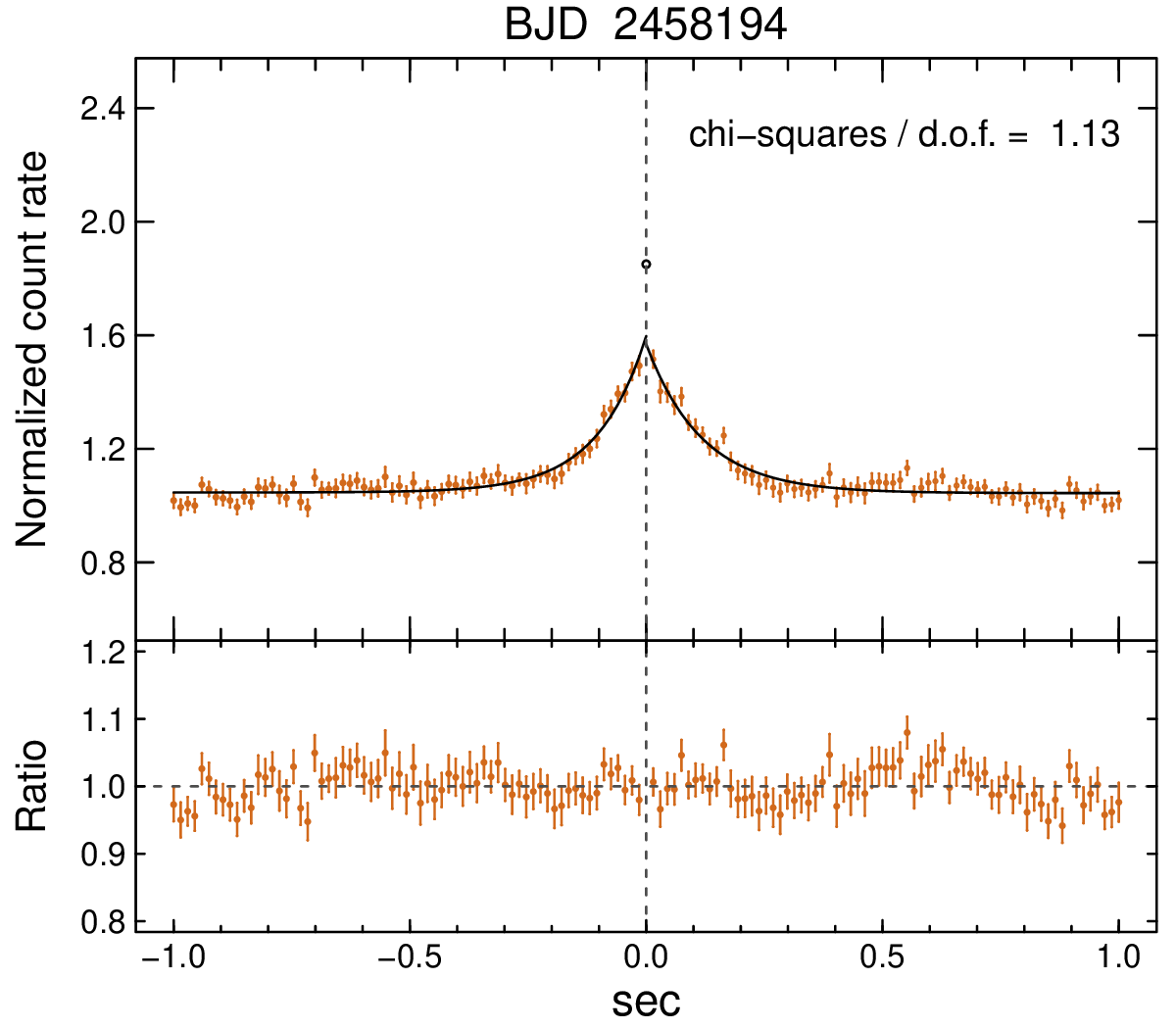}
\end{minipage}
\begin{minipage}{0.325\hsize}
\FigureFile(50mm, 50mm){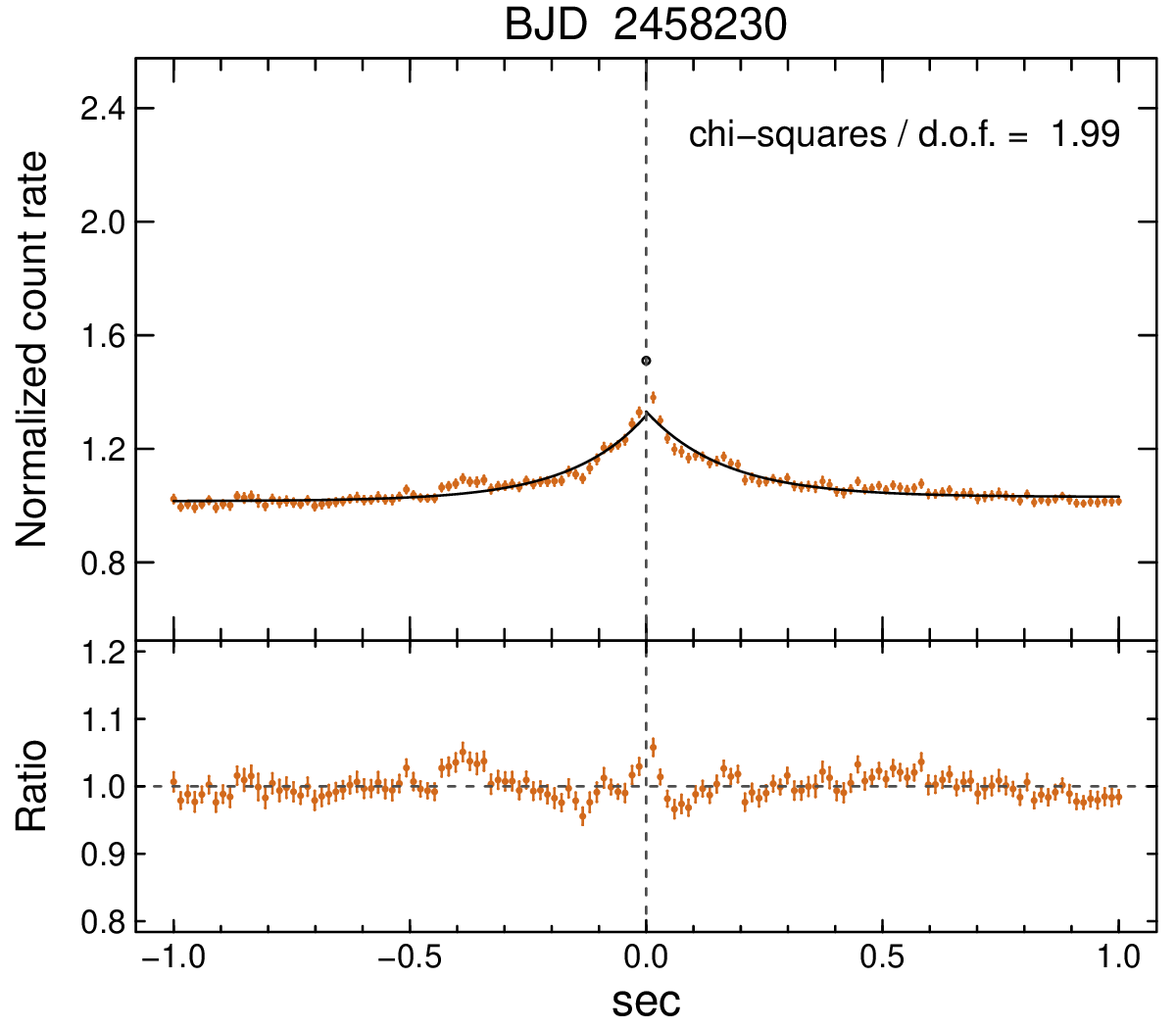}
\end{minipage}
\begin{minipage}{0.325\hsize}
\FigureFile(50mm, 50mm){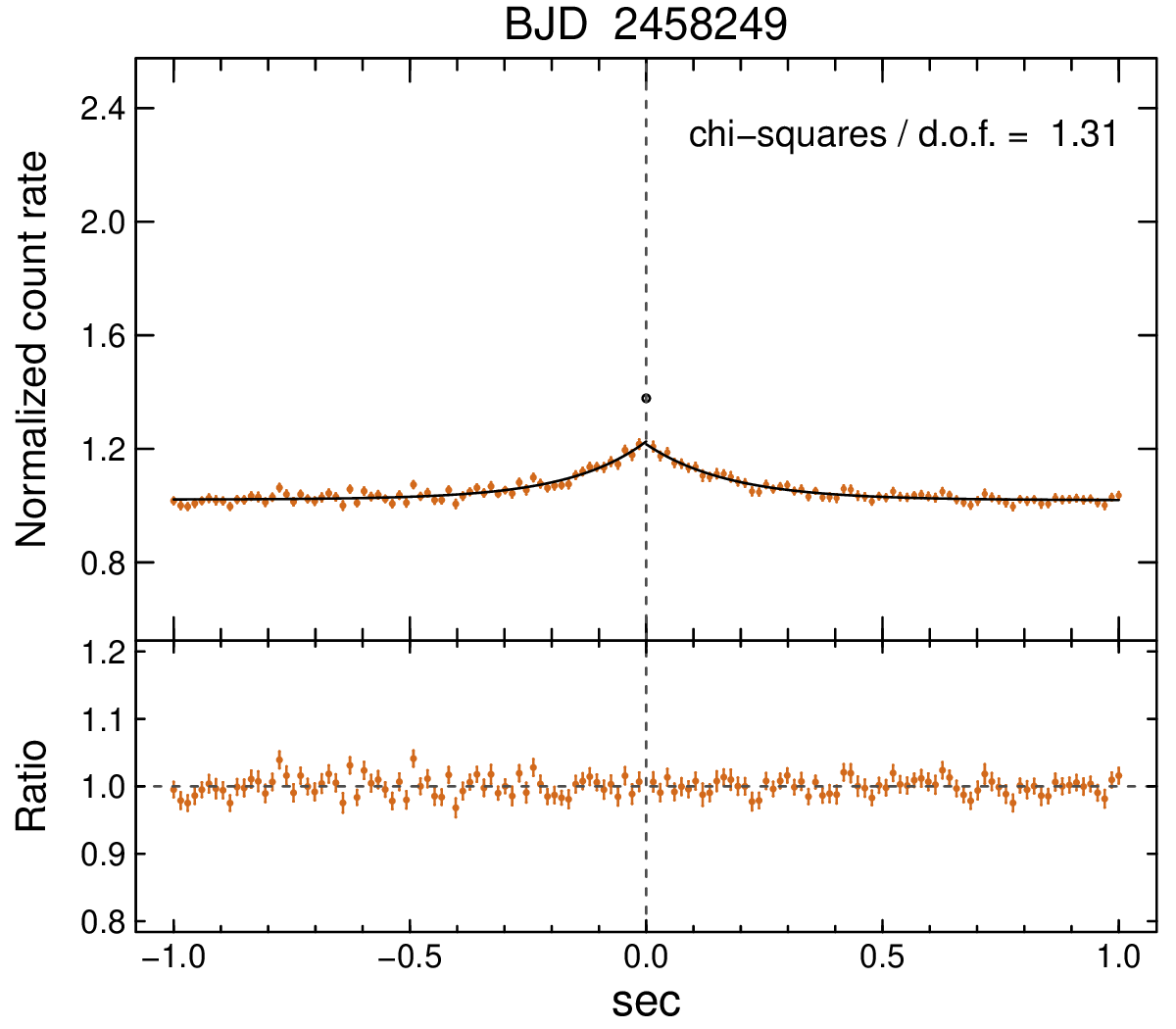}
\end{minipage}
\\
\begin{minipage}{0.325\hsize}
\FigureFile(50mm, 50mm){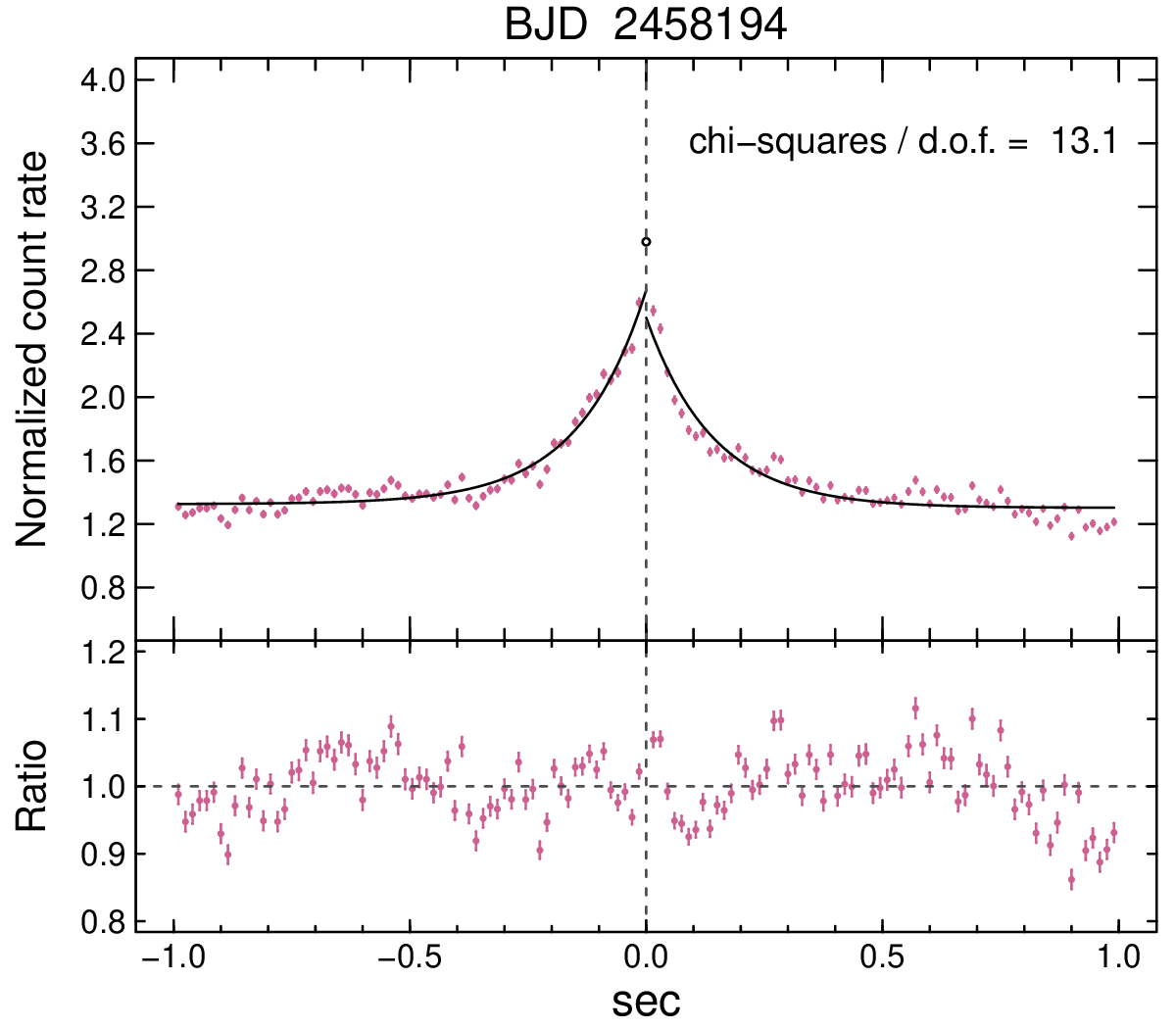}
\end{minipage}
\begin{minipage}{0.325\hsize}
\FigureFile(50mm, 50mm){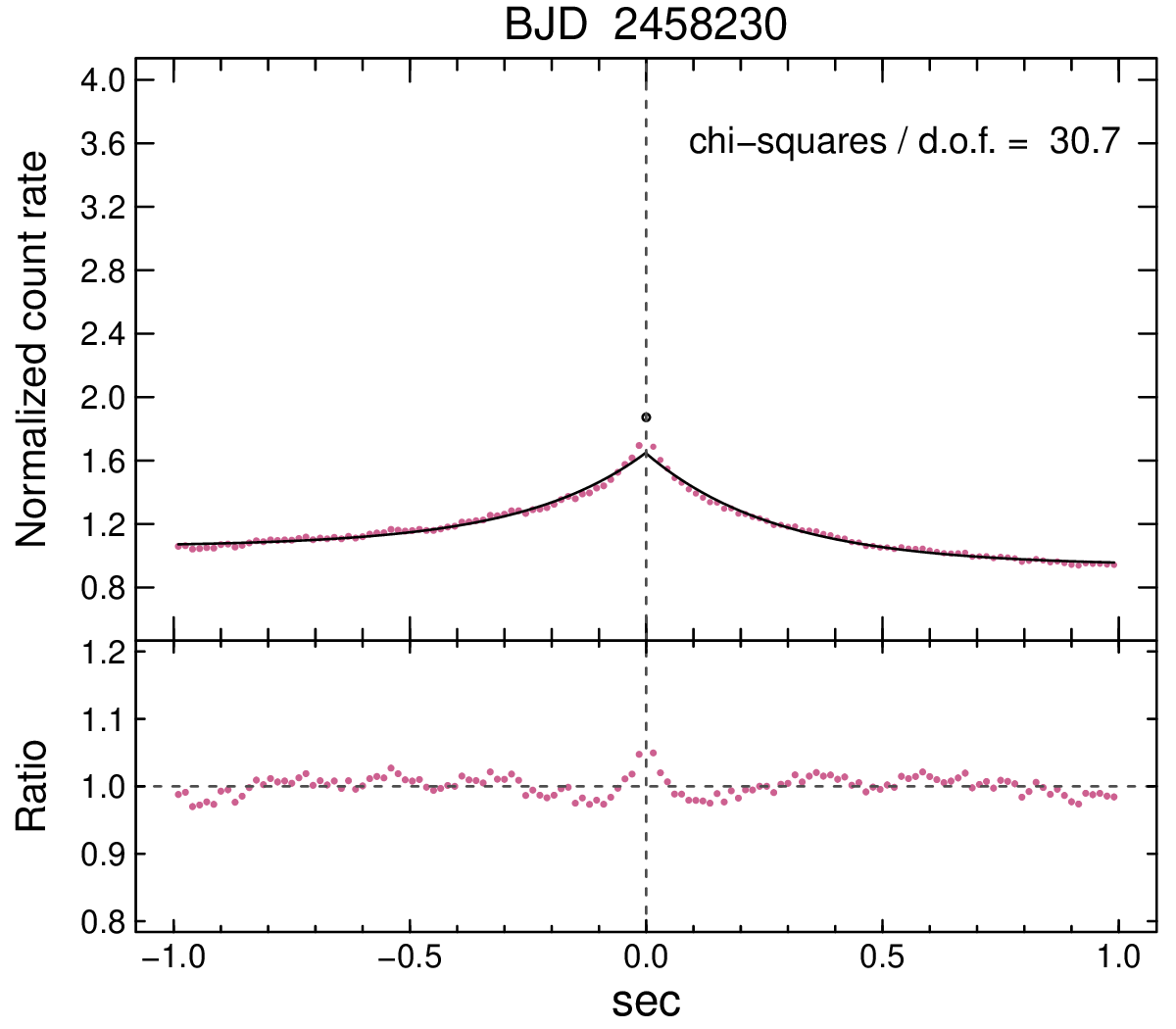}
\end{minipage}
\begin{minipage}{0.325\hsize}
\FigureFile(50mm, 50mm){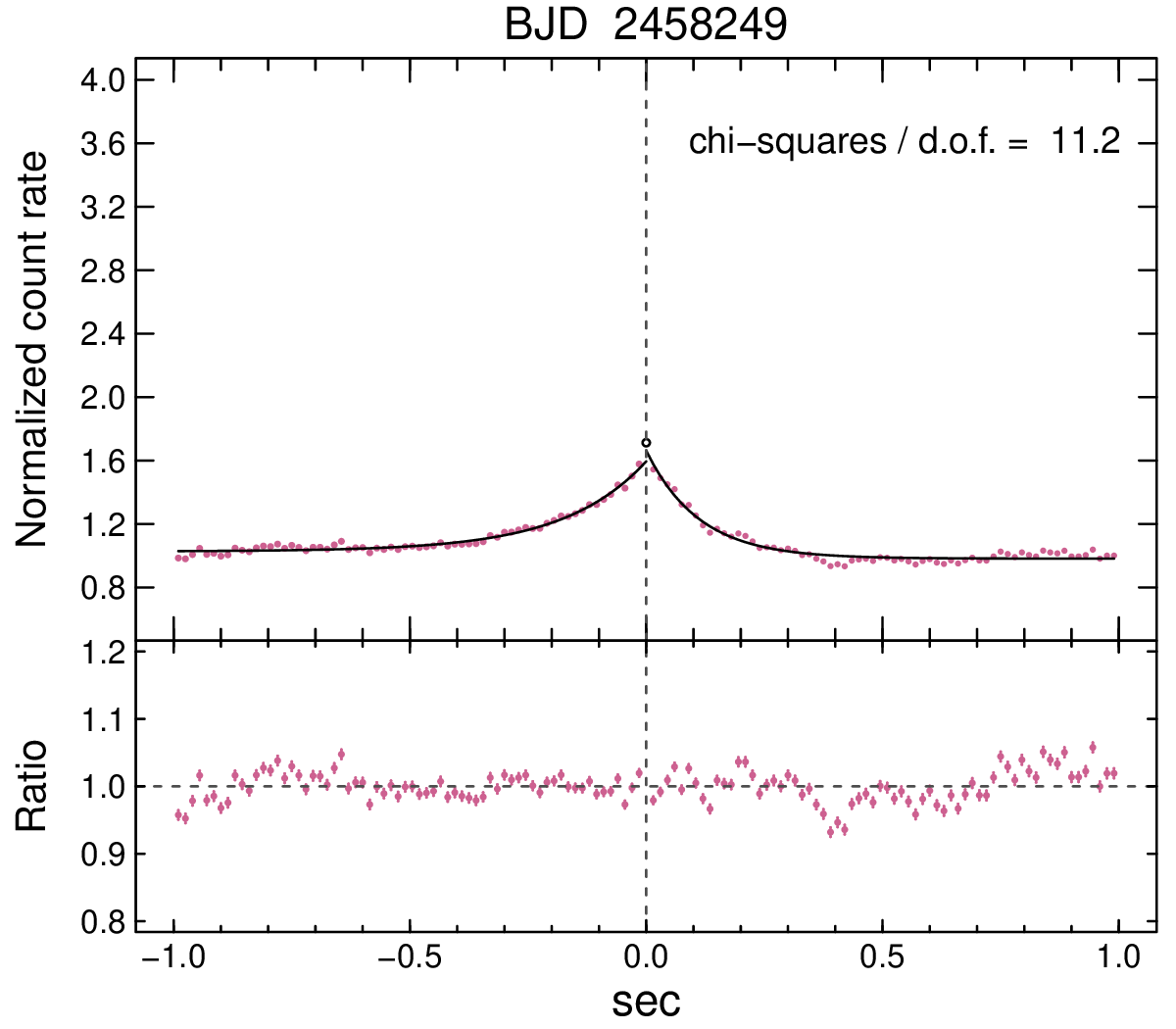}
\end{minipage}
\\
\begin{minipage}{0.325\hsize}
\FigureFile(50mm, 50mm){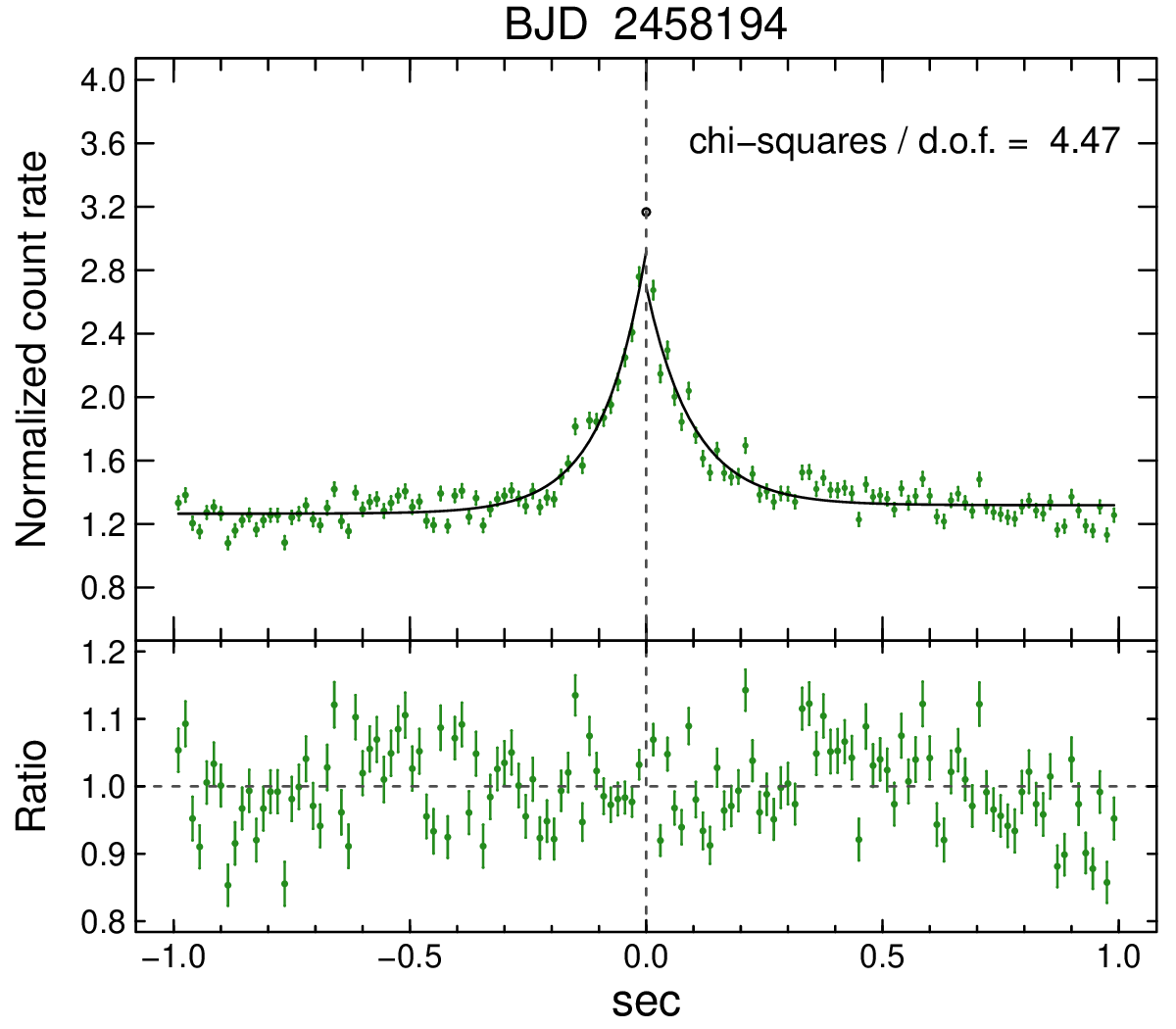}
\end{minipage}
\begin{minipage}{0.325\hsize}
\FigureFile(50mm, 50mm){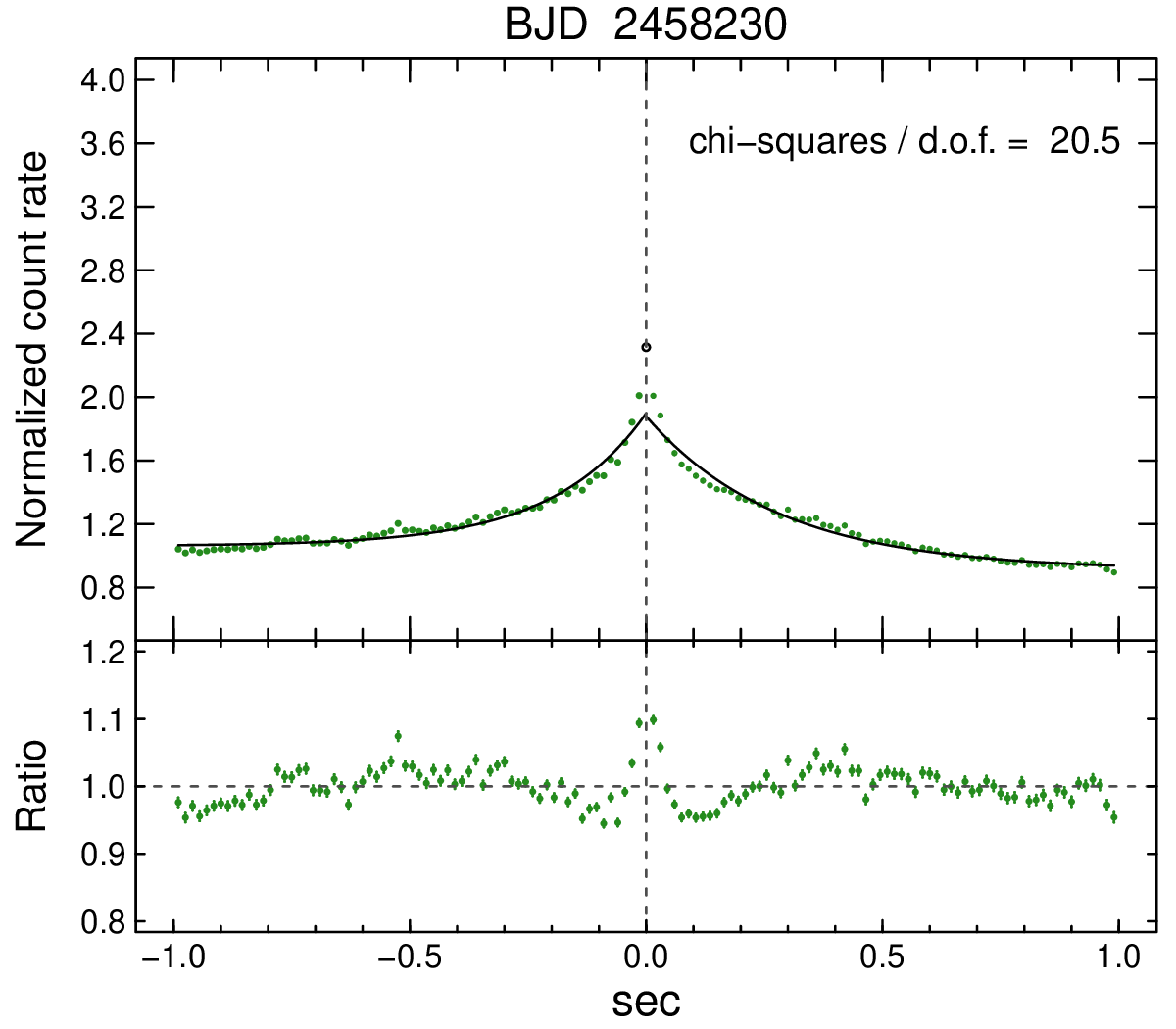}
\end{minipage}
\begin{minipage}{0.325\hsize}
\FigureFile(50mm, 50mm){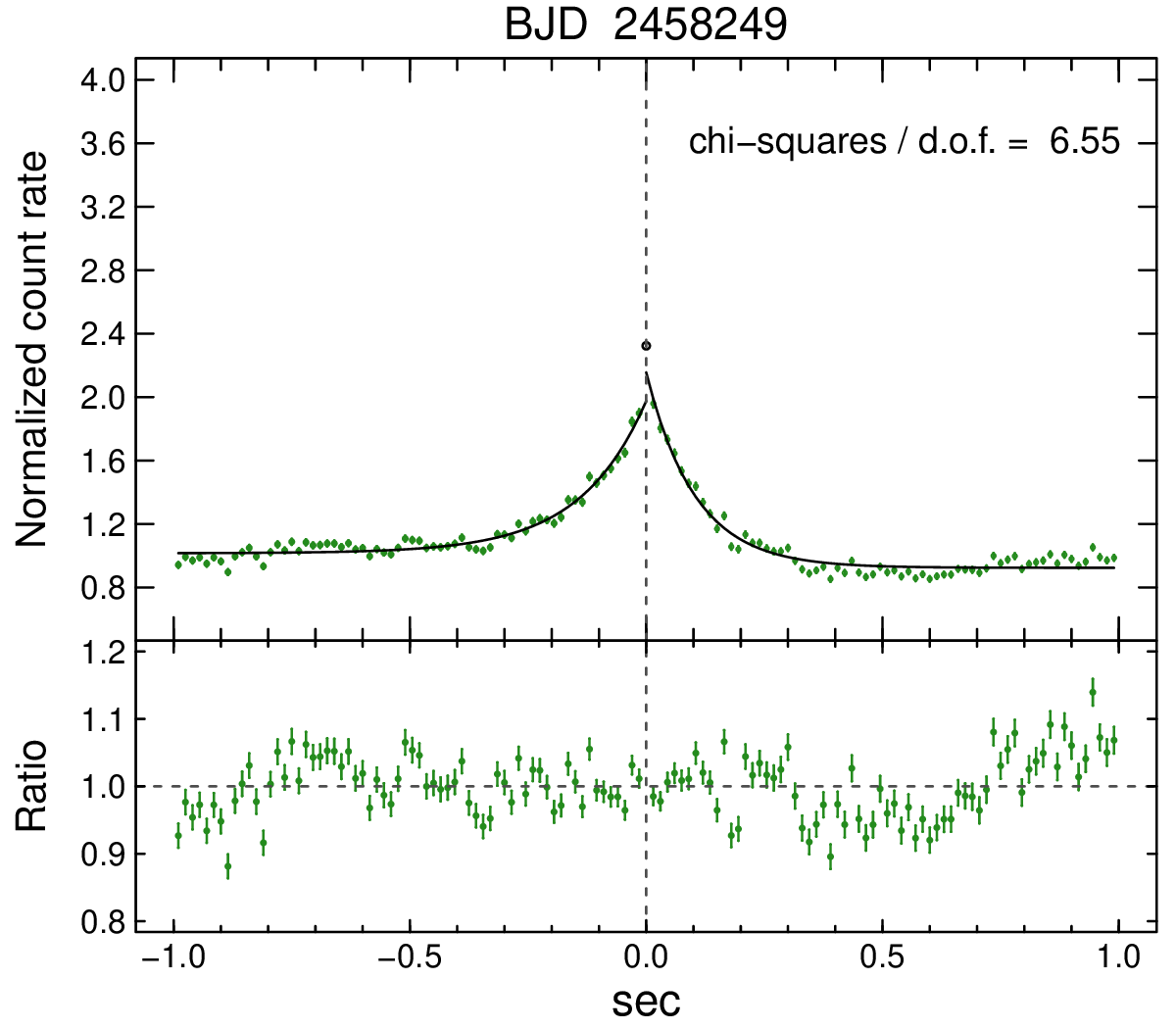}
\end{minipage}
\end{center}
\caption{
Results of modeling of optical and X-ray shots.
In each figure, the upper panel shows the best-fit model alongside the data, and the lower panel shows the ratio of the data to the model.
The orange, pink, and green crosses represent the data of optical, soft X-ray, and hard X-ray shots.
The black line represents the best-fit model.
The reduced $\chi^2$ value is given in the upper panel of each figure.
}
\label{fitting-res}
\end{figure*}
%\fi

\textcolor{black}{
We give the representative ones of the figures for the goodness-of-fit of our modeling in Figure \ref{fitting-res}.
The exponential function in equation (\ref{exp-func}) well reproduced optical shots and most X-ray shots.
However, X-ray shots during H2 and the first half of H3 deviated from this function.
This was imprinted on the steep spectral hardening near the shot peak (see the panel of the hardness ratio in Figures \ref{optical-xray-shots} and \ref{xray-shots}).
%In these time periods, the shot duration may be overestimated or underestimated.
In these time periods, the shot duration may not be accurately estimated.
Hereafter, we excluded the data in these time periods when estimating the shot duration.
Past works suggested that only one exponential function sometimes did not reproduce the shot profile well 
%and that another function was more suitable 
\citep{neg94cygx1shot,sas17w2r1926,dob19mvlyr}.
%The spectrum kept hard, and another component generating the long tail was likely superposed to the narrow shot.
Another component generating the long tail was likely superposed to the narrow shot.
The exponential function is only intended to obtain a quantitative measure of the spread of the shot and has no physical background.
In this paper, we focus on roughly examining the evolution of the shot duration, and examining what kind of functions well fit the shot profile in H2 and H3 and retrying shot selections are our future work.}

\begin{figure*}[htb]
\begin{center}
\FigureFile(120mm, 50mm){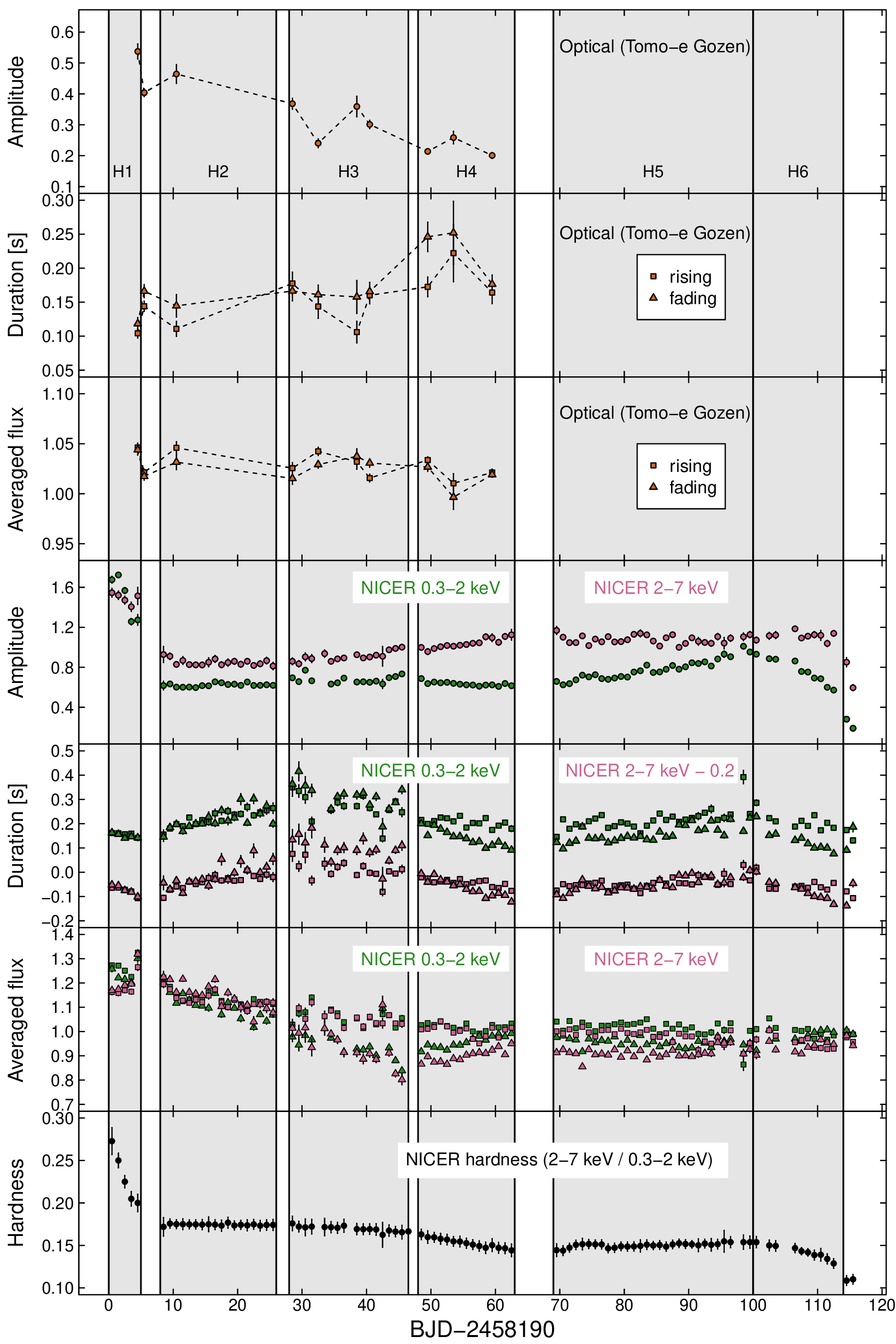}
\end{center}
\caption{
Properties of optical and X-ray shots during the low/hard state of the 2018 outburst in MAXI J1820$+$070.
The top, second, and third panels show the normalized amplitude, duration, average flux outside of flares of optical shots, respectively.
The fourth, fifth, and sixth panels show the normalized amplitude, duration, and average flux outside of flares of X-ray shots, respectively. 
The bottom panel shows the average ratio of the 2--7~keV count rate to the 0.3--2~keV count rate.
Here, the green and pink colors represent the 0.3--2 keV data and the 2--7 keV data, respectively.
The normalized amplitude is calculated by $(F_1 + F_2)/2$ in equation (\ref{exp-func}).
The rectangle and triangle in the second and fifth panels represent $\tau_1$ and $\tau_2$ in equation (\ref{exp-func}), respectively.
The rectangle and triangle in the third and sixth panels represent $c_1$ and $c_2$ in equation (\ref{exp-func}), respectively.
In the fifth panel, the data denoted by the pink color are vertically shifted by $-$0.2~s for visibility.
We plot the time periods of five spectral states classified by the hardness-intensity diagram in Figure \ref{q-diagram} with grey boxes, and write their names in the top panel.
}
\label{shot-property}
\end{figure*}

The shot property is summarized in Figure \ref{shot-property}.
The average value of $(F_1 + F_2)/2$ for optical shots was 0.33, and that for X-ray shots was 0.68.
The amplitude of the optical shot was about half of the X-ray one.
\textcolor{black}{The sum of $\tau_1$ and $\tau_2$ for optical shots was 0.29--0.47~s and that for X-ray shots was 0.29--0.74~s.
The average values of $\tau_1+\tau_2$ for optical and X-ray shots were 0.33~s and 0.40~s, respectively.
In H1 and H4, the duration of optical shots was comparable with that of X-ray shots.
In H3, the optical shot was narrower than the X-ray shot by $\sim$250~ms on average.
Here, we estimated these values for X-ray shots before BJD 2458250 in the 0.3--7~keV band.}
Both optical and X-ray shots were more or less asymmetric to $t_p$.
%The optical shot seems to be symmetric to $t_p$.
The asymmetry to $t_p$ for X-ray shots was more prominent and seems to have developed since H3.
The optical flux at $t \gtrsim 0.5$ became higher than the X-ray one, and the difference between $c_1$ and $c_2$ became larger since then (see the top panels of Figure \ref{optical-xray-shots}).

The asymmetric profile of X-ray shots came from the original flare shape.
Figure E2 in the supplementary information represents four examples of X-ray flares selected via the process of our shot analysis on BJD 2458247.
The rapid decline was imprinted in these profiles.
In the low/hard state, quasi-periodic oscillations (QPOs) having periods of 1--100~s were observed at optical and X-ray wavelengths \citep{bui19j1820,sti20j1820,tho22j1820,gao23j1820,shu23j1820}.
The QPO period was reported to be shorter than 10~s after around BJD 2458230.
We found that wavy modulations having periods close to the QPO period developed in the 10-s length shot profile (see Figure E4 in the supplementary information).
The flux of the shot tail became lower than the average flux in H3 and increased in H4.
The wavy modulation was more prominent in the harder band, and the hardness also showed weak modulations.
These QPOs might dilute the shot profile.
We will investigate the relation between the shot and QPO phases in future work.
%and it is suitable to discuss the asymmetry of the shot profile after that.

\subsubsection{Time evolution of X-ray shots} \label{sec:xray-shots}

\begin{figure*}[htb]
\begin{center}
\begin{minipage}{0.325\hsize}
\FigureFile(50mm, 50mm){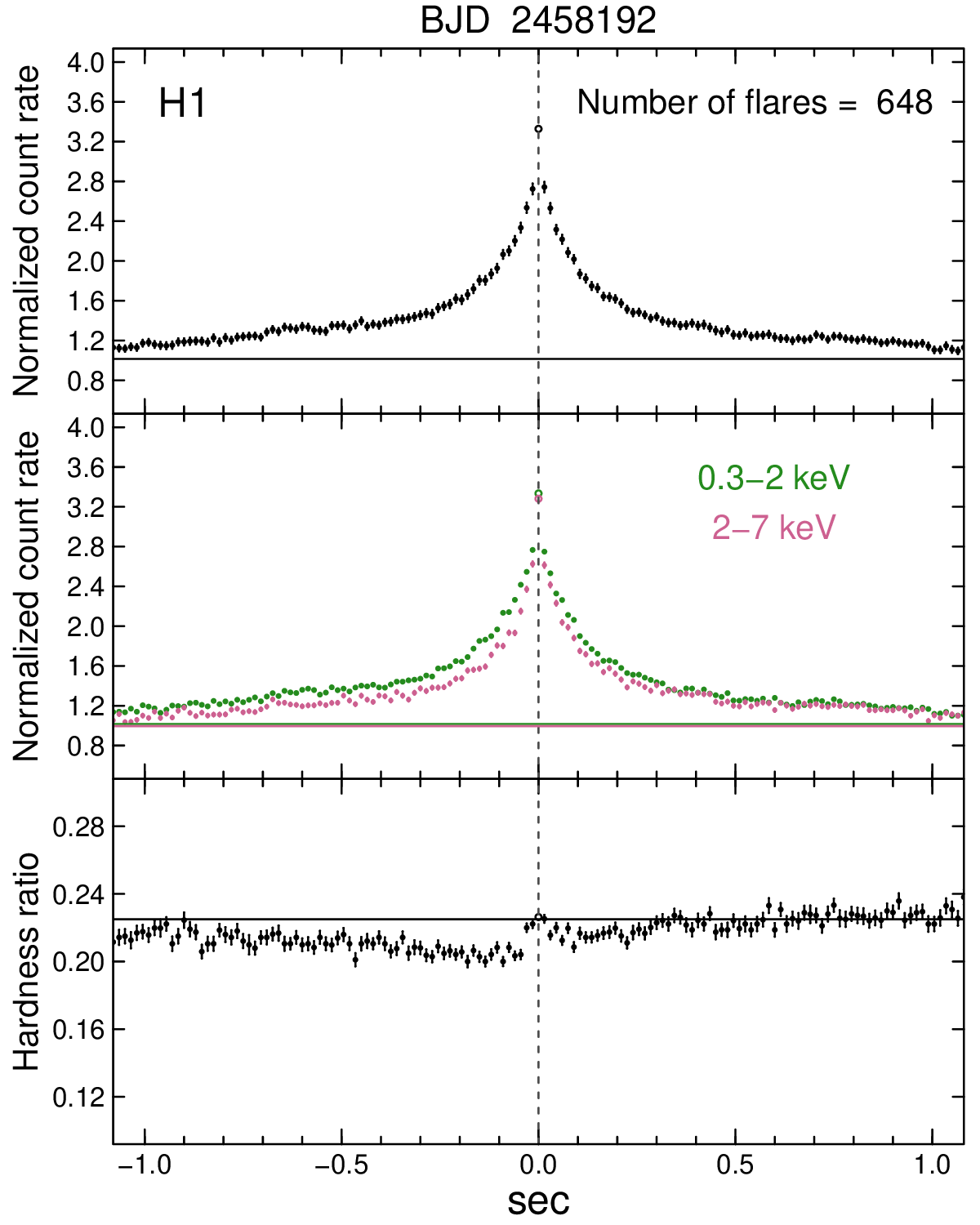}
\end{minipage}
\begin{minipage}{0.325\hsize}
\FigureFile(50mm, 50mm){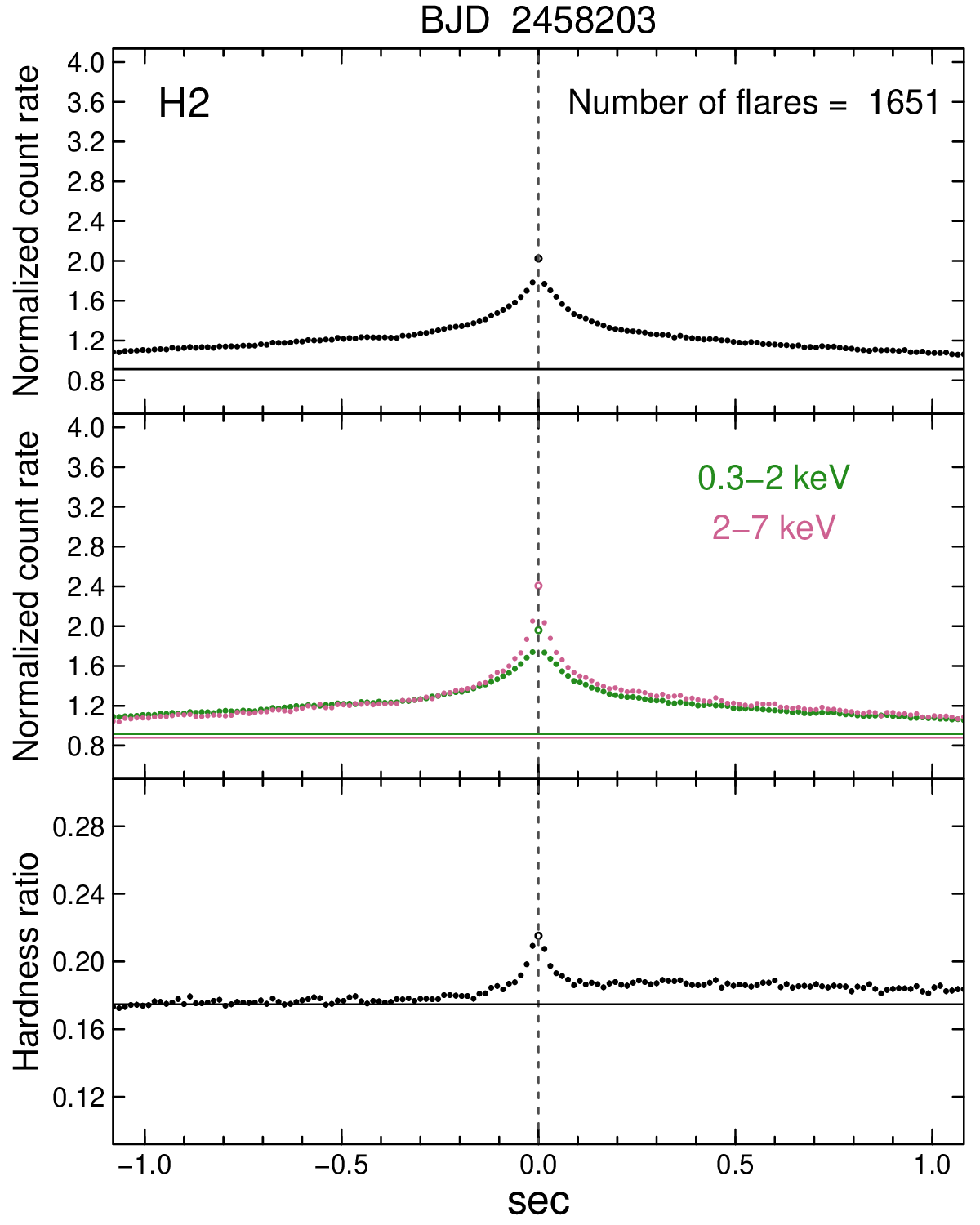}
\end{minipage}
\begin{minipage}{0.325\hsize}
\FigureFile(50mm, 50mm){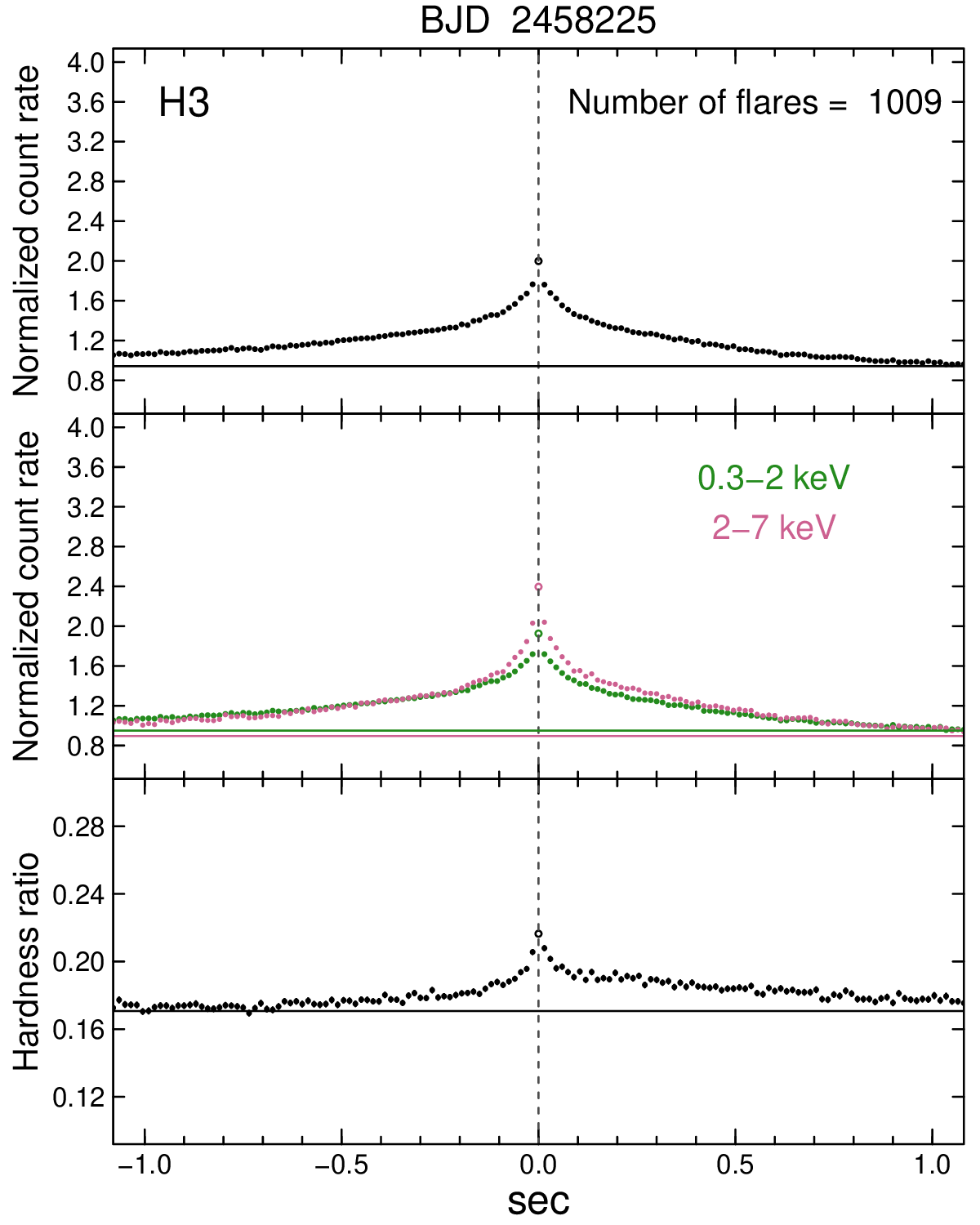}
\end{minipage}
\\
\begin{minipage}{0.325\hsize}
\FigureFile(50mm, 50mm){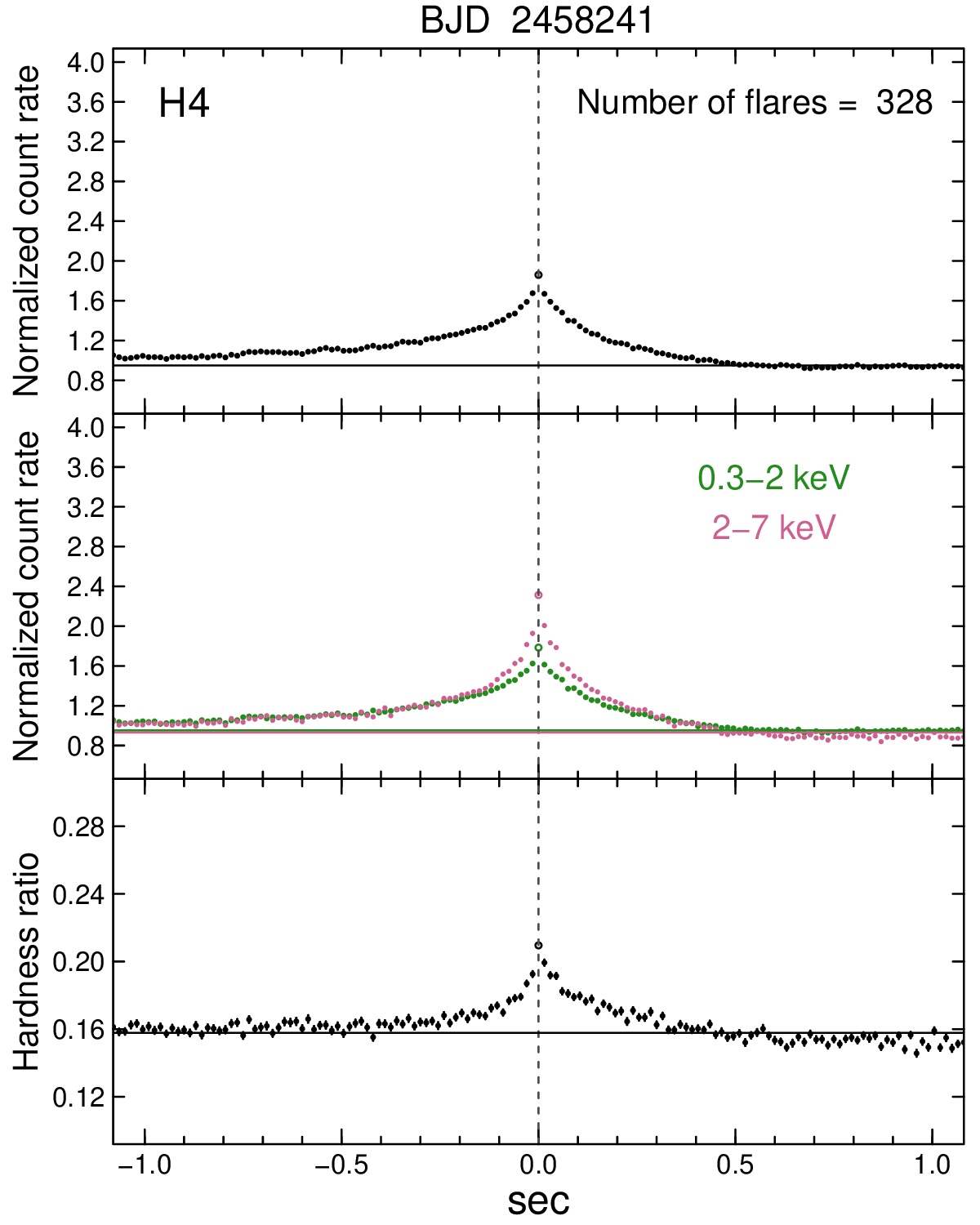}
\end{minipage}
\begin{minipage}{0.325\hsize}
\FigureFile(50mm, 50mm){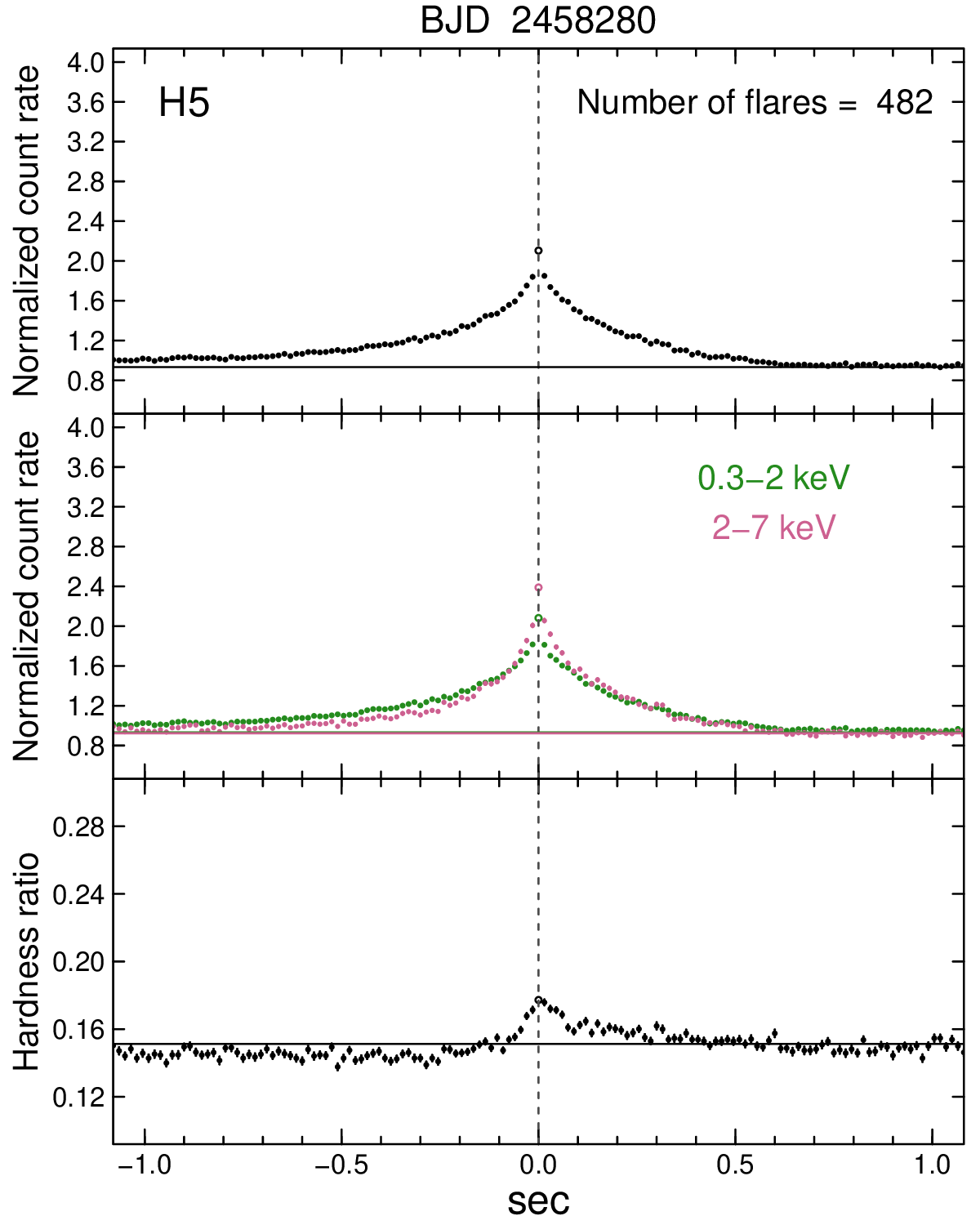}
\end{minipage}
\begin{minipage}{0.325\hsize}
\FigureFile(50mm, 50mm){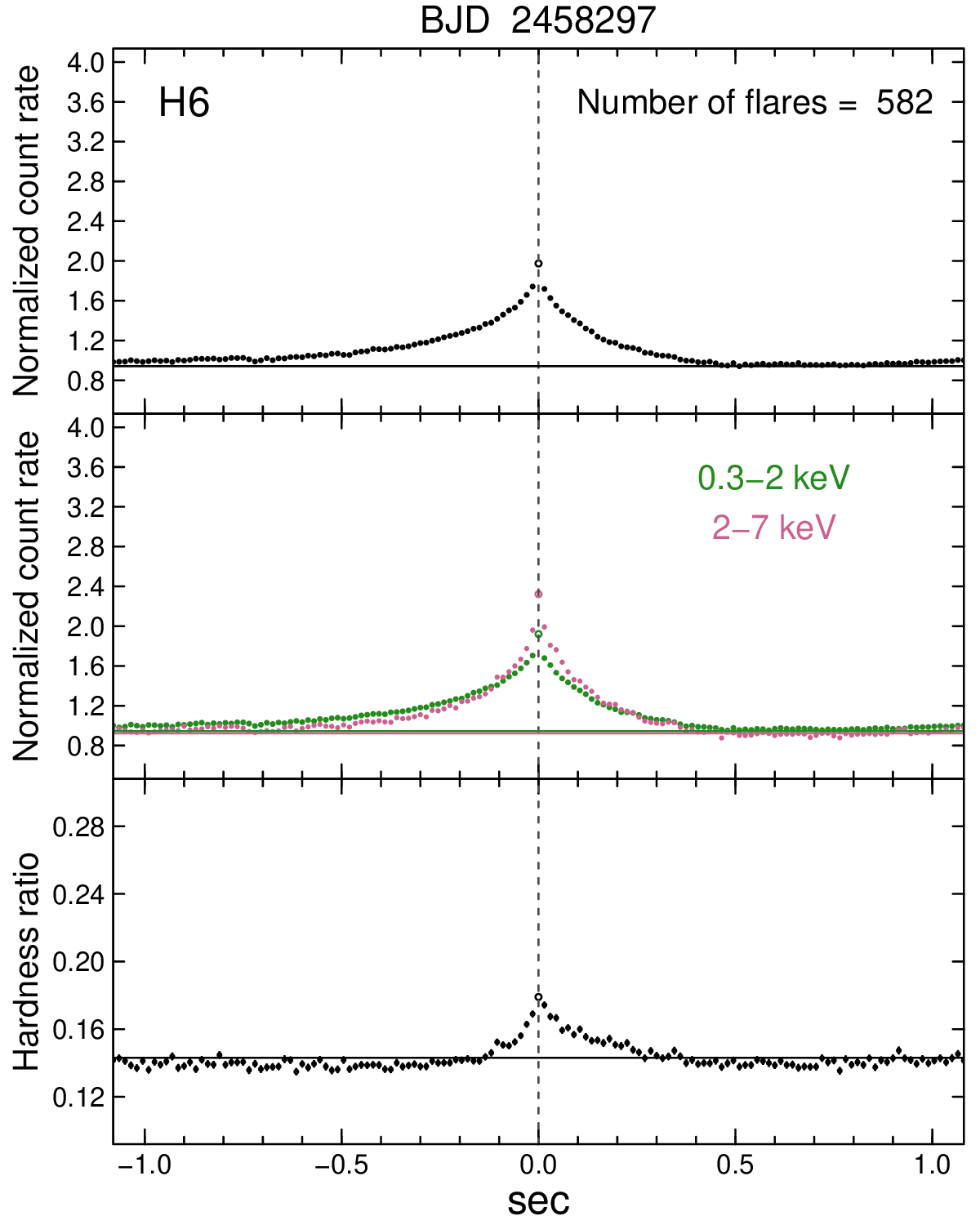}
\end{minipage}
\end{center}
\caption{
Representative X-ray shot profiles averaged per date with the evolution of the hardness ratio in the phases of H1, H2, H3, H4, H5, and H6 during the low/hard state of the 2018 outburst in MAXI J1820$+$070.
We give the number of flares superposed for one shot profile is given in the top panel.
The top panel displays the X-ray shot profile in the 0.3--7 keV band (black).
The middle panel shows the soft X-ray shot in the 0.3--2 keV band (green) and the hard X-ray shot in the 2--7 keV band (pink).
The bottom panel is the 2--7~keV / 0.3--2~keV hardness ratio of the X-ray shot.
The horizontal lines in the top and middle panels represent the average flux level outside of flares at $|t| > 2.5$~s.
The horizontal line in the bottom panel represents the average hardness ratio.
}
\label{xray-shots}
\end{figure*}

As mentioned in section \ref{sec:opt-xray-shots}, the profile of X-ray shots was time-varying.
The time variations of the amplitude and duration seem to have been correlated with the change in phases between H1--H6.
%We also found that the hardness also changed with time.
In addition to Figure \ref{shot-property}, we prepared Figure \ref{xray-shots} that displays the representative shot profile in each phase and its hardness to check the evolution of the actual profile and the hardness.
We also give the shot profile and its hardness every $\sim$10~d in Figure E3 in the supplementary information to explore their evolution in detail.

During H1, the amplitude rapidly decreased.
The amplitude of soft shots was comparable with that of hard shots.
The duration also decreased.
%Since H2, the hard shot became larger than the soft shot.
During H2, the soft shot was smaller than the hard shot, and their amplitudes were almost constant.
\textcolor{black}{The duration would become longer.}
During H3, the amplitude of hard shots gradually increased, while that of soft shots stayed at the same level as that during H2.
\textcolor{black}{The shot duration seemed to be long} and decreased with time.
The profile became asymmetric to $t_{\rm p}$, a faster rise and slower decay (see the difference between the rectangles and triangles in the fifth panel of Figure \ref{shot-property}).
Past studies showed that the shot profile in black-hole binaries, cataclysmic variables, and active galactic nuclei tended to be a faster rise and slower decay \citep{dob19mvlyr}.
The luminosity dip after the fading was remarkable (see the triangles in the sixth panel of the same figure).
During H4, the hard shot became even larger, while the amplitude of soft shots was almost constant.
The shot duration became shorter.
The shot profile became slower rise and faster decay.
The difference in timescales between the rising and fading parts became bigger and bigger, in particular for soft shots.
The luminosity dip after the fading became less remarkable.
During H5, the amplitude of hard shots was almost constant, but soft shots became larger. 
Its amplitude was close to that of hard shots at the end of this phase.
The shot duration gradually increased.
The shot profile was slow rise and rapid decay.
During H6, the soft shot rapidly faded, though the hard shot still remained.
The shot duration decreased.
The shot profile kept a slow rise and rapid decay.

\textcolor{black}{The average value of the shot duration for soft and hard shots was 0.37~s and 0.31~s, respectively.
The hard shot was $\sim$20\% narrower than the soft shot on average.}
Since the light curve was normalized before the shot analysis, the value of $(c_1 + c_2)/2$ should be around unity.
However, this value was as high as 1.3 and gradually approached unity (see also the sixth panel of Figure \ref{shot-property}).
Actually, the flux at the edge of the shot was higher than the average flux level outside of the shot till H2.
The narrow shot may be superposed by another modulation whose duration was much longer than the shot duration, which might be QPOs, as mentioned in section \ref{sec:opt-xray-shots}.

The change in the hardness within the one-shot profile is classified into three stages. 
First, we see a small decrease just before the rising part, which was prominent during H1 and H5.
Second, the hardness became harder coinciding with the sharp flaring, which would correspond to the narrower hard shot than the soft shot as mentioned above.
Such behavior was not found in previous shot analyses for another black-hole binary Cyg X-1 \citep{neg94cygx1shot,yam13cygx1}.
Third, the hardness did not return to the same level after the fading and stayed at a slightly harder level for $\gtrsim$1~s, which was confirmed in all time periods except H4.

\section{Discussion}

\subsection{Possible sources of X-ray rapid variability} \label{sec:discuss-xray-shot}

The shot amplitude was the highest at the onset of the outburst, during H1 (see the fourth panel of Figure \ref{shot-property}), at which the Comptonization component was dominant in soft and hard X-rays \citep{shi19j1820}.
Also, X-ray flares suddenly disappeared at the X-ray spectral state transition from the low/hard state to the intermediate state after H6 (see the fourth panel of Figure \ref{shot-property}).
It was reported that the multi-temperature disk overshined the Comptonization component just after the state transition \citep{shi19j1820,bui21j1820}.
It is, therefore, naturally considered that the X-ray rapid variability originated from the advection-dominated accretion flow (ADAF)-like region in the vicinity of the black hole, from which X-ray photons are emitted by inverse Compton scattering \citep{nar94ADAF}.
In fact, it was confirmed by past spectral analyses that the Comptonization component became brighter and its electron temperature became lower during X-ray flares, which suggests that the properties of the Comptonization component changed in flaring events \citep{shi18j1820}.

As pointed out by \citet{neg94cygx1shot} for Cyg X-1, two different timescales are involved in the X-ray rapid variability also for MAXI J1820$+$070.
One is the timescale of the spectral hardening after the fading of the shot, which continued for $\gtrsim$1~s (see Figure \ref{xray-shots} and Figure E4 in the supplementary information).
The other one is the timescales of the rising and fading parts of the shot, which was $\sim$0.2~s (see the fifth panel of Figure \ref{shot-property}).
Also, the increase and decrease of the hardness synchronized with the flaring event occurred on this short timescale.
The gas in the ADAF accretes onto the black hole with the velocity of $v_r (R) = \alpha (H/R)^2 v_\phi (R)$.
Here, $R$, $\alpha$, $H$, and $v_\phi$ are the radial distance from the central black hole, the viscosity parameter, the scale height, and the rotation velocity, respectively.
Although it is unclear whether the hot flow extended vertically or radially during the low/hard state of the 2018 outburst in MAXI J1820$+$070, it has spacial extension \citep{bui19j1820,kar19j1820,xu20j1820,zdz21j1820}.
For example, it was reported that the radius of the ADAF-like flow was $\sim$50~$r_{\rm g}$, where the gravitational radius $r_{\rm g}$ is defined as $G M_1 / c^2$ \citep{xu20j1820,zdz21j1820}.
The viscous timescale is $\sim$1~s at $R \sim 50~r_{\rm g}$.
Here, we applied $H/R \sim 1$ and $\alpha \sim 0.01$ \citep{nar97ADAFBHeventhorizon}, which is consistent with the observed longer timescale.
However, some physical processes with the dynamical or thermal timescale would be necessary for explaining the steep shot and accompanying hardening.
Even in the case of the vertically-extended hot flow, its height might exceed 10~$r_{\rm g}$ at some points \citep{kar19j1820}.

For instance, magnetic reconnection in the innermost region of the black-hole accretion flow occurs on timescales of $1/v_{\rm A} \sim 1/\alpha^{1/3} \Omega$, where $v_{\rm A}$ is the Alfv\'{e}n velocity \citep{gal79corona}.
This timescale is equivalent to the dynamical timescale of the disk.
Random mass accretion from the disk towards the black hole powers magnetic flares \citep{min94BHfluctuation,pou99magnetic}.
\citet{mac03globalMHD} performed three-dimensional global resistive magnetohydrodynamic (MHD) simulations and proposed that the accretion of dense gas blobs deposits the energy to the accretion flow, which triggers the magnetic reconnection.
Hot electrons and protons are produced by a series of magnetic reconnections, which amplify hard X-ray emission.
It is expected that the peak of soft X-ray flares, which would be determined by the timing of the accretion of gas blobs, precedes that of hard X-ray flares and that harder flares are narrower than softer flares.
The liberated magnetic energy is converted to heat, and the gas is gradually cooled, which produces spectral hardening lasting for a while after the fading of flares.

A sequence of the processes described in the previous paragraph could explain the three stages of the hardness variations described in section \ref{sec:xray-shots} and the short duration of X-ray shots, though we did not find the delay of the peak of hard shots to that of soft shots.
The delay time would be too short to be detected (see Figure 2(c) in \cite{pou99magnetic}).
Although the expected sharp increase of the hardness at the peak time of X-ray shots was not confirmed in Cyg X-1 \citep{neg94cygx1shot,neg01cygx1shot,yam13cygx1}, we succeeded in detecting it.
\citet{yam13cygx1} reported that the shot was observable at more than 100~keV.
Our detected spectral hardening at the steep flare may indicate a sudden increase in the electron temperature due to the abrupt release of magnetic energy.
As discussed in \citet{kaw22j1820}, if we introduce other types of accretion flows with shorter viscous timescales, which were proposed by \citet{nar12sane} and \citet{mar18jed}, the short shot duration could be explained.
It is common for magnetic activity to be prominent in these flows.
Also, only fluctuations of mass accretion rates could not explain the observed shot amplitude of more than 60\% \citep{man96ADAFfluctuation}.

Suppose the primary source of X-ray rapid variability is the magnetic activity in the hot flow. 
In that case, it is natural that the amplitude of the soft X-ray shot would be lower than that of the hard X-ray shot during H2--H6 (see the fourth panel of Figure \ref{shot-property}), when the emission from the optically-thick accretion disk contributed to the soft X-ray flux \citep{shi19j1820,xu20j1820,zdz21j1820}.
%The significant contribution of the disk emission to the soft X-ray flux was pointed out as the reason for low shot amplitudes in the energy band softer than $\sim$2~keV for Cyg X-1 \citep{bha22cygx1}.
The shot amplitudes in the energy bands softer than $\sim$2~keV were also low for Cyg X-1 \citep{bha22cygx1}.
The lower amplitude in the softer shot is consistent with the power spectrum analysis performed by \citet{axe21j1820}, in which the power of soft X-ray variations with timescales of less than 1~s was lower than that of hard X-ray variations.
The duration of the soft X-ray shot was $\sim$20\% longer than that of the hard X-ray shot.
This wider profile may originate from the process for launching flares as discussed in the previous paragraph, or may come from the reprocessing of hard X-ray photons at the optically-thick disk.
In the latter case, the time delay of soft X-ray variations relative to hard X-ray variations, which is called the soft lag, would be observed.
In fact, the soft lag with timescales of $\sim$1~ms was reported by \citet{kar19j1820,wan21j1820,dem21j1820,kaw22j1820,oma23j1820}.
Since we used the NICER light curves with a 0.015-s bin, we were not able to detect this kind of very short time lag.
The time delay of hard X-ray variations relative to soft X-ray variations, which is called the hard lag, has also been detected in past works.
The hard lag was a few seconds, and the variability timescale was larger than 10~s \citep{kaw22j1820,oma23j1820}.
This hard lag may be associated with the spectral hardening lasting for $\gtrsim$1~s after the fading from the shot peak.
According to \citet{oma23j1820}, the hard lag became shorter during the low/hard state, and the duration of the spectral hardening after the shot tail, which we detected, also became shorter with time (see Figure \ref{xray-shots}).

The asymmetry of the shot profile developed since H3 (see the fifth panel of Figure \ref{xray-shots}).
This evolution of the asymmetry may be caused by superposed QPO signals, as mentioned in section \ref{sec:opt-xray-shots}, or may reflect some physical mechanisms for producing flares with slow or rapid fading.
\citet{mac03globalMHD} expected that hard X-ray flares show exponential decay since their sources are instantaneous acceleration of electrons by magnetic reconnection (M.~Machida, private communication). 
On the other hand, the density disturbance in the hot flow reproduces flares having an almost symmetric profile to the peak time but showing a slightly deeper dip at the fading stage \citep{man96ADAFfluctuation}.
Also, \citet{neg02cygx1} suggested the temporary depletion of the accreting gas after the flaring, which may be the cause of the luminosity dip after the fading from the shot.
The hard X-ray shot profile would be determined by the magnetic event, while the soft X-ray shot would be composed of fluctuations of the accretion rate, the energy dissipation by magnetic reconnection, and the reprocessed emission.
The soft X-ray shot profile would easily deviate from the symmetric one in comparison with the profile of hard X-ray shots.
Our observations showed that the slower decay was more remarkable in hard X-ray shots and that the faster decay was more remarkable in soft X-ray shots (see the fifth panel of Figure \ref{shot-property}), and the above-mentioned physical process may be imprinted in the shot profile.

\subsection{Radiation mechanism of optical rapid variability} \label{sec:discuss-optical-shot}

%The radiation mechanism of optical rapid variability has been discussed for a long time.
As possible radiation mechanisms of optical rapid variability, the reprocessing of X-ray photons at the low-temperature outer accretion disk and the synchrotron radiation in the hot flow and/or the jet launching plasma have been proposed \citep{kan01j1118varcorrelation,vel11sscmodel,kim16v404cyg,gan16v404cyg}.
If the reprocessing is dominant, the optical shot is expected to become a smeared shape in comparison with the X-ray shot.
\textcolor{black}{However, the optical shot was less spread than the X-ray shot (see Figures \ref{optical-xray-shots} and \ref{shot-property})}.
Our results support the synchrotron radiation rather than the X-ray reprocessing at least for rapid variations with timescales of a few hundreds of milliseconds.
The correlation analysis between optical and X-ray variations during the 2018 outburst of MAXI J1820$+$070 showed optical lags with timescales of $\sim$0.2~s and a preceding dip at negative lags in the cross-correlation functions (CCFs), which also suggests that the synchrotron emission from internal shocks within jet plasmas and/or the synchrotron self-Compton emission in the hot flow \citep{pai19j1820,kaj19j1820,pai21j1820,tho22j1820}.
This observational feature was also detected in XTE J1118$+$480 (= KV UMa), another black-hole binary, and a similar interpretation was proposed \citep{kan01j1118varcorrelation,spr02j1118}.
We note that the NICER and Tomo-e Gozen light curves analyzed in section \ref{sec:opt-xray-shots} were not completely simultaneous observational data.
We, therefore, could not test whether the reported optical lag was detectable or not by our shot analyses.
The average duration of the optical shot was 0.32~s as estimated in section \ref{sec:opt-xray-shots}, which implies that the size of the emission region was $\sim$7900~$r_{\rm g}$ under the assumption that the optical flare occurs on the dynamical timescale.

The entire time evolution in the NICER hardness vs.~the optical flux diagram was similar to that of the hardness-intensity diagram during H2--H4, though the optical flux was more variable than the X-ray one in the same phases (see Figure \ref{q-diagram}).
This suggests that the synchrotron emission from the hot flow significantly contributed to the optical continuum emission during the low/hard state of the 2018 outburst in MAXI J1820$+$070.
On the other hand, some physical processes, except for the magnetic activity in the hot flow, would influence the optical rapid variability.
If the source of the optical rapid variability is the same as that of the X-ray rapid variability, the time evolution of the optical and X-ray shot properties should be correlated with each other.
However, the amplitude of the optical shot decreased during H2, while that of the X-ray shot was almost constant during this period.
Also, the duration of the optical shot increased during H3, while the X-ray shot showed the opposite behavior.
The simultaneously observed optical and X-ray variability seemed not to necessarily correlate with each other in all time periods \citep{pai19j1820}, which suggests that some optical flares were independent of X-ray flares.
We consider that the jet emission contributed to the optical variability, as discussed in the previous paragraph.
The multi-wavelength spectral analysis performed by \citet{shi18j1820} showed that the jet emission did not significantly contribute to the X-ray flux.
On the other hand, the optical flux in the low/hard state was much higher than expected from the reprocessed disk \citep{van94visualLMXB} according to \citet{shi19j1820}.

The optical shot seems to have faded at the end of H4 since its amplitude became lower and lower (see Figure \ref{optical-xray-shots} and the top panel of Figure \ref{shot-property}).
The appearance of superhumps in H5 was reported by \citet{nii21j1820}.
This implies that the emission from the outer disk became dominant at optical wavelengths after H5.
There is a possibility that the rapid variability became less conspicuous by the radiation from the accretion disk.

\subsection{Evolution of properties of rapid variability} \label{sec:discuss-shot-evolution}

In the hardness-intensity diagram, we found six different phases in the low/hard state of the 2018 outburst in MAXI J1820$+$070 (see the left panel of Figure \ref{q-diagram}), and the properties of X-ray shots seem to have changed with the evolution of these phases (see Figure \ref{shot-property}).
This may represent the time evolution of the hot Comptonization component with strong magnetic fields.
\citet{zdz21j1820} and \citet{kaw22j1820} suggested that the hard X-ray emission from this object originated from two different components.
The detailed geometry and how many components consist of the hot flow are still unclear; however, the harder Comptonization component would be the closest to the black hole, and the softer Comptonization component would be located outside of the harder one.
A similar idea was proposed for Cyg X-1 \citep{yam13bcygx1}.
Since the energy spectra of these two components were similar to each other, it is difficult to distinguish them only by spectral analyses.
Instead, the evolution of the shot property may give the key information for understanding the time evolution of multiple components of hot Comptonization flows.

For example, the duration of X-ray shots increased during H2 and H5 and decreased during H3, H4, and H6.
If the size of the emission region becomes larger, the duration becomes larger.
If only a part of the hot flow having strong magnetic fields is the source of the rapid X-ray variability, the increase and decrease in the shot duration will represent the expansion and shrinkage of the source region of the rapid X-ray variability.
It is suggested by past X-ray timing and spectral analyses that the hot flow contracted in the vertical or horizontal axis during H3--H4 \citep{kar19j1820,bui19j1820,dem21j1820,zdz21j1820}.
The decrease in the shot duration during H3--H4 may be consistent with these results.
Our shot analyses could not determine whether the disk extended to the ISCO or was truncated.
The combination with spectral analyses, which are our future work, may address this problem.
On the other hand, the change in the duration of optical shots was not synchronized with that of X-ray shots.
As discussed in the previous section, the jet emission was likely dominant at optical wavelengths till H4. 
In this case, the gradual increase in the optical shot duration during H3--H4 may indicate the expansion of the jet base.
This interpretation seems to be consistent with that in \citet{wan20j1820}.
They showed that the delay time of hard X-ray variations to soft X-ray variations increased till H4 and decreased after that, and argued that this result indicates the expansion and shrinkage of jet plasmas.

The amplitude of X-ray shots also increased and decreased during the low/hard state, and the behavior was different between soft and hard shots.
The disk emission became dominant in the 0.3--2~keV band after H2.
The change in the contribution of the hot flow to the soft X-ray flux would lead to the variation in the amplitude of soft X-ray shots.
During H3 and H4, the gap in the amplitude between hard shots and soft shots became larger (see the fourth panel of Figure \ref{shot-property}).
The spectrum became softer during these phases (see the bottom panel of the same figure), which suggests the more significant contribution of the disk emission with time, as discussed in section \ref{sec:discuss-xray-shot}.
On the other hand, the amplitude difference between hard and soft shots became smaller during H5.
\citet{wan20j1820} suggested that this phase was similar to the failed outburst.
The disk emission would be less dominant in the 0.3--2~keV band during this phase.
Not only the amplitude but also the change in the hardness of X-ray shots during H5 seem to have been similar to those during H1 (see Figure \ref{xray-shots}).

We need to consider complex physical mechanisms in discussing the time evolution of the amplitude of hard X-ray shots since the mass accretion rate and the magnetic energy do not directly converted into the hard X-ray emission.
During H3 and H4, the amplitude of hard X-ray shots gradually increased.
More electrons might be accelerated by a series of magnetic reconnections during these phases, which results in more hard X-ray photons via the inverse Compton scattering.

Here, we summarize our proposed picture of the time evolution of the entire accretion flow during the low/hard state in the 2018 outburst of MAXI J1820$+$070.
During H1, the rapid increase in mass accretion rates to the hot inner flow abruptly enhanced X-ray and optical emission.
The emission from the hard Comptonization component with strong magnetic fields was dominant over wide X-ray energy ranges.
\textcolor{black}{During H2, the disk emission became dominant in soft X-rays.}
The jet was launched, which mainly affected the optical emission.
During H3--H4, the source flow for causing X-ray rapid variability might contract, and some components, except for that flow, developed even in hard X-rays.
The jet base expanded with time.
\textcolor{black}{The total size of multiple Comptonization components would become smaller.}
During H5, the jet plasma shrank, and the tidal dissipation at the outer disk became the main source of the optical continuum.
The hard Comptonization component was dominant again in soft X-rays with the decrease in the temperature of the disk.
During H6, the inner accretion disk became hotter again, and the hard Comptonization component kept contributing to hard X-ray emission.

\section{Summary} \label{sec:summary}

We classified the low/hard state during the 2018 outburst of MAXI J1820$+$070 into six phases by using the optical and X-ray overall light curves, and performed shot analyses by using high-time-cadence X-ray and optical light curves taken by NICER and Tomo-e Gozen.
Our main results and their interpretations are as follows:
\begin{itemize}
        
\item \textcolor{black}{The optical shot was less spread than the X-ray shot.}
This implies that the radiation mechanism of optical shots was the synchrotron emission at the hot inner accretion flow and/or the jet plasma (see sections \ref{sec:opt-xray-shots} and \ref{sec:discuss-optical-shot}).

\item The amplitude of X-ray shots was maximized at the onset of the outburst, and that of hard X-ray shots was higher than that of soft X-ray shots after that.
At the X-ray spectral transition from the low/hard state to the intermediate state, the shot rapidly faded.
The soft X-ray shot was wider than the hard X-ray shot.
The timescale of shots was $\sim$0.2~s.
These results suggest that the main source of the rapid X-ray variability was the magnetic activity in the hot, optically-thin, and geometrically-thick accretion flow in the vicinity of the black hole (see sections \ref{sec:xray-shots} and \ref{sec:discuss-xray-shot}).

\item We detected the spectral hardening synchronized with the steep shot for the first time.
We also found that the spectrum kept harder after the fading from the shot for $\gtrsim$1~s.
These features could be explained by the abrupt energy release of a series of magnetic reconnections, which are induced by the accretion of gas blobs, and the gradual cooling of the heated gas (see sections \ref{sec:xray-shots} and \ref{sec:discuss-xray-shot}).

\item The time evolution in the amplitude and duration differed between optical and X-ray shots.
This implies that optical emission was influenced by another component except for the hot inner accretion flow, which may be jet ejections (see sections \ref{sec:opt-xray-shots} and \ref{sec:discuss-optical-shot}).

\item The X-ray shot showed different behavior with different spectral phases.
The amplitude and duration increased or decreased in each phase.
If the shot duration reflects the size of the main source of rapid X-ray variability and the shot amplitude is determined by the contribution of the disk emission, our results may provide key information for understanding the evolution of the hot inner accretion flow, the outer cool accretion disk, and the jet emitting plasma (see sections \ref{sec:xray-shots} and \ref{sec:discuss-shot-evolution}).

\end{itemize}

In this paper, we focus on investigating the entire evolution of the properties of X-ray and optical shots.
The detailed spectral analyses in each shot profile, the exploration of the relation between shots and QPOs, and the cross-correlation analysis between soft X-ray, hard X-ray, and optical shot profiles are our future work.

\section*{Acknowledgements}

This research was partially supported by the Optical
and Infrared Synergetic Telescopes for Education and Research
(OISTER) program funded by the Ministry of Education, Culture,
Sports, Science and Technology (MEXT) of Japan.
This research is also supported in part by the Research Center for the Early Universe (RESCEU) of the School of Science at the University of Tokyo.
We are thankful to many amateur observers for providing a lot of data used in this research.  
M.~Kimura acknowledges Mami Machida, who commented on the X-ray emission by magnetic reconnection.
This work was financially supported by Japan Society for the Promotion of Science Grants-in-Aid for Scientific Research (KAKENHI) Grant Numbers JP20K22374 (MK), JP21K13970 (MK), JP21H04491 (SS, MK, HN), JP21K03620 (HN), JP18H05223 (SS), JP17H06363 (SS), JP16H06341 (SS), JP16H02158 (SS), JP26247074 (SS), JP25103502 (SS), JP20H01942 (SS), JP18H04575 (SS), and JP24K00673 (WI).
\textcolor{black}{We thank the anonymous referee for his or her insightful comments.}

%\appendix

%\section*{1.~~Details of the modifications of our numerical code}

%\fi

\bibliography{pasjadd,/Users/mariko/cvs2}
\bibliographystyle{pasjtest1}

\end{document}